\documentclass[sigconf, nonacm=true]{acmart}
\usepackage{booktabs} 
\usepackage{subcaption} 
\usepackage{amssymb} 
\usepackage{xcolor, colortbl} 
\usepackage{array,multirow,graphicx} 
\usepackage{makecell} 
\setcopyright{none}

\usepackage{todonotes}

\acmYear{}
\acmConference[ACSAC '18]{2018 Annual Computer Security Applications Conference}{December 3--7, 2018}{San Juan, PR, USA}
\acmDOI{}
\acmISBN{}

\makeatletter
\def\@copyrightspace{\relax}
\makeatother
\definecolor{mygray}{gray}{0.85}
\begin{document}
\title{Tracking Users across the Web via TLS Session Resumption}

\author{Erik Sy}
\affiliation{%
  \institution{University of Hamburg}
}

\author{Christian Burkert}
\affiliation{%
  \institution{University of Hamburg}
}

\author{Hannes Federrath}
\affiliation{%
  \institution{University of Hamburg}}

\author{Mathias Fischer}
\affiliation{%
  \institution{University of Hamburg}}
\renewcommand{\shortauthors}{Sy et al.}

\begin{abstract}
User tracking on the Internet can come in various forms, e.g., via cookies or by fingerprinting web browsers.
A technique that got less attention so far is user tracking based on TLS and specifically based on the TLS session resumption mechanism.
To the best of our knowledge, we are the first that investigate the applicability of TLS session resumption for user tracking. For that, we evaluated the configuration of $48$ popular browsers and one million of the most popular websites. 
Moreover, we present a so-called prolongation attack, which allows extending the tracking period beyond the lifetime of the session resumption mechanism. To show that under the observed browser configurations tracking via TLS session resumptions is feasible, we also looked into DNS data to understand the longest consecutive tracking period for a user by a particular website. 
Our results indicate that with the standard setting of the session resumption lifetime in many current browsers, the average user can be tracked for up to eight days. With a session resumption lifetime of seven days, as recommended upper limit in the draft for TLS version 1.3, $65$\% of all users in our dataset can be tracked permanently.
\end{abstract}

%
%

\ccsdesc[500]{Security and privacy}
\ccsdesc[500]{Security and privacy~Browser security}
\ccsdesc[500]{Security and privacy~Privacy protections}

\keywords{Session IDs, Session Tickets, PSK Identity, Tracking Period, Browser Measurement}

\maketitle

\section{Introduction}

User tracking via HTTP cookies and browser fingerprinting is a reality~\cite{englehardt2016online,papadopoulos2018cookie,gomez2018hiding}. 
Tracking mechanisms are commonly used to observe conversions, namely whether an advertisement on website \textit{A} leads to a desired user activity on website \textit{B}.
Herrmann et al.~\cite{herrmann2013behavior} revealed that temporary tracking mechanisms can be used to identify users based on their characteristic browsing patterns over longer time periods. They found that $85,4\%$ of users can be identified based on their browsing behaviour if the temporary tracking mechanism lasts up to 24 hours. 
Also, the creation of long-term browsing profiles is possible, if a tracker can observe a large share of a user's browsing activity. Big players like Google and Facebook leverage the wide-spread use of their advertising networks and social plugins to track users across websites and gain detailed user profiles. 

However, as users are increasingly aware of the privacy threat from tracking, they use privacy-friendly browsers, private browsing modes, and browser extensions to restrict tracking practices such as HTTP cookies.
Browser fingerprinting got more difficult, as trackers can hardly distinguish the fingerprints of mobile browsers. They are often not as unique as their counterparts on desktop systems~\cite{gomez2018hiding, laperdrix2016beauty}. 
Tracking based on IP addresses is restricted because of NAT that causes users to share public IP addresses and it cannot track devices across different networks. 
As a result, trackers have an increased interest in additional methods for regaining the visibility on the browsing habits of users. The result is a race of arms between trackers as well as privacy-aware users and browser vendors. 

One novel tracking technique could be based on TLS session resumption, which allows abbreviating TLS handshakes by leveraging key material exchanged in an earlier TLS session. Thus, it introduces a possibility to link two TLS sessions. 
However, continuous user tracking via TLS session resumption is only possible as long as the browser is not restarted, because this clears the TLS cache. Especially mobile devices are \textit{always on} and seldomly restarted.
Finally, the feasibility of user tracking via TLS session resumption depends on the TLS configuration of both server and browser, as well as on the user's browsing behaviour. It is unknown so far whether this approach is feasible for user tracking in real-world scenarios. 

\begin{figure*}[htpb]
\begin{minipage}{.33\linewidth}
    \vspace{-3.78 mm}
    \includegraphics[width=\linewidth]{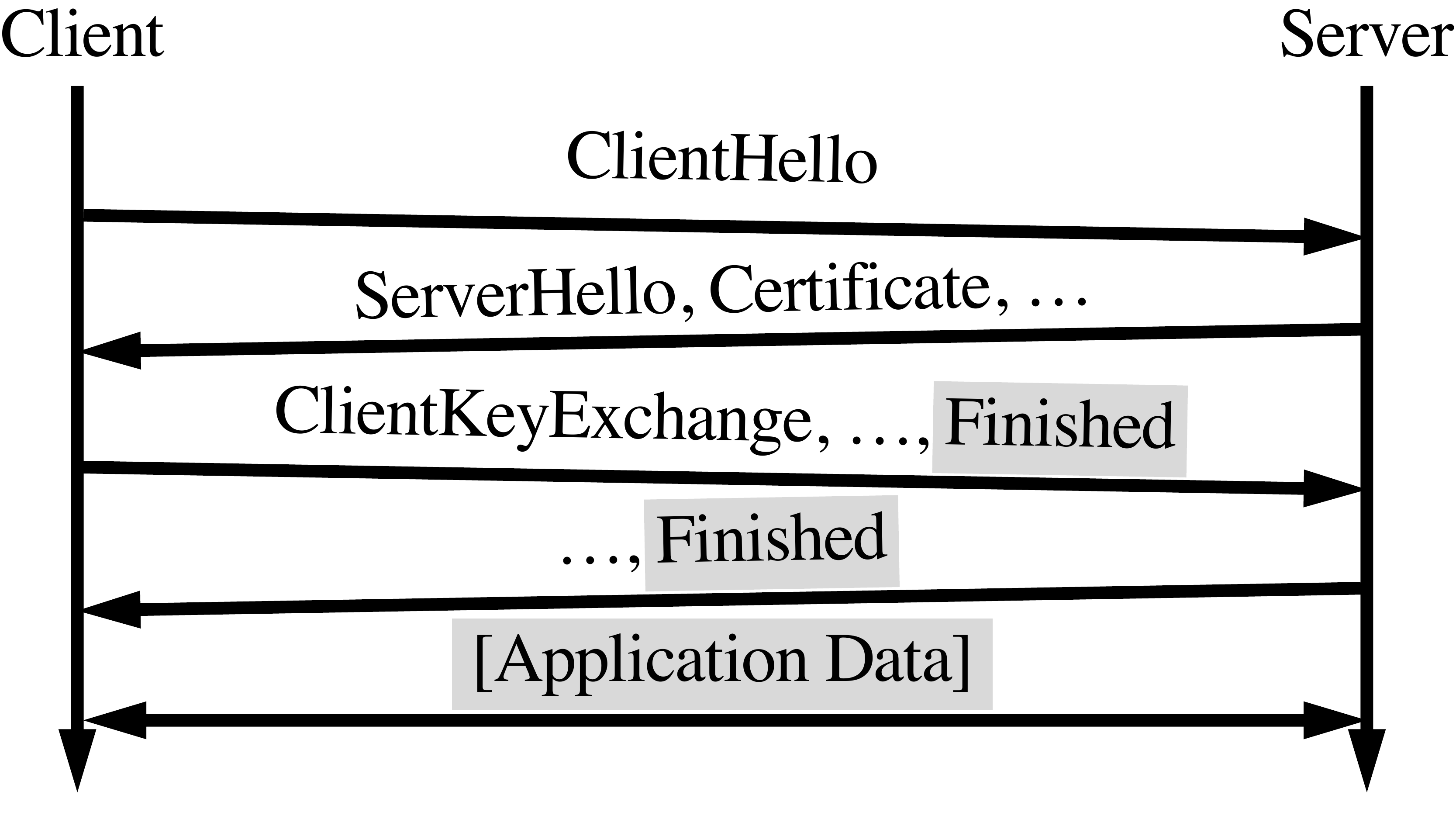}
    \vspace{-10 mm}
    \caption*{a) Full Handshake}
    \label{fig:tls12}
\end{minipage}
\hfill
\begin{minipage}{.33\linewidth}
    \includegraphics[width=\linewidth]{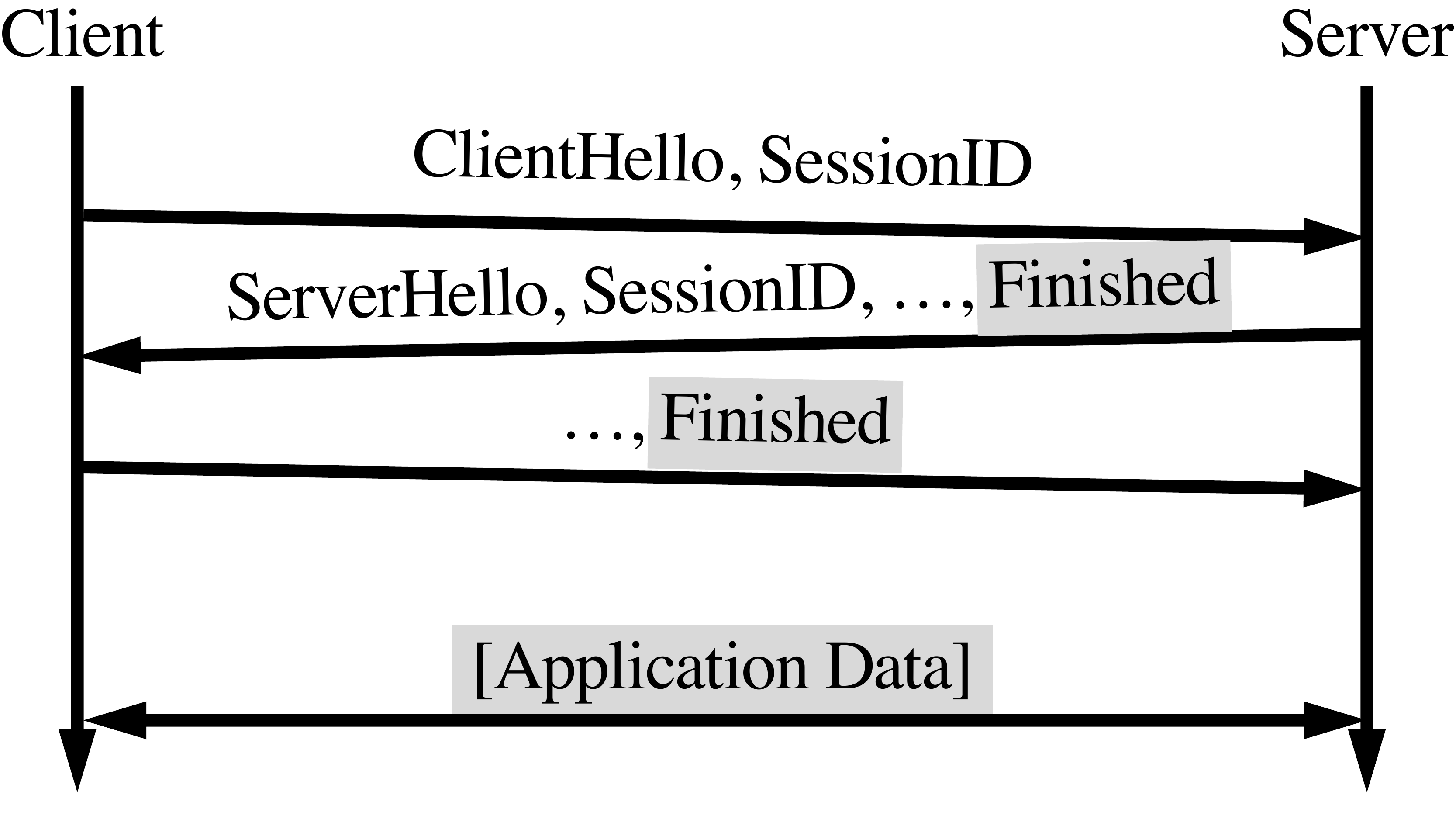}
     \vspace{-10 mm}
    \caption*{b) Session Resumption with \\ Session ID}
    \label{fig:tls12_id}
\end{minipage}
\hfill
\begin{minipage}{.33\linewidth}
    \includegraphics[width=\linewidth]{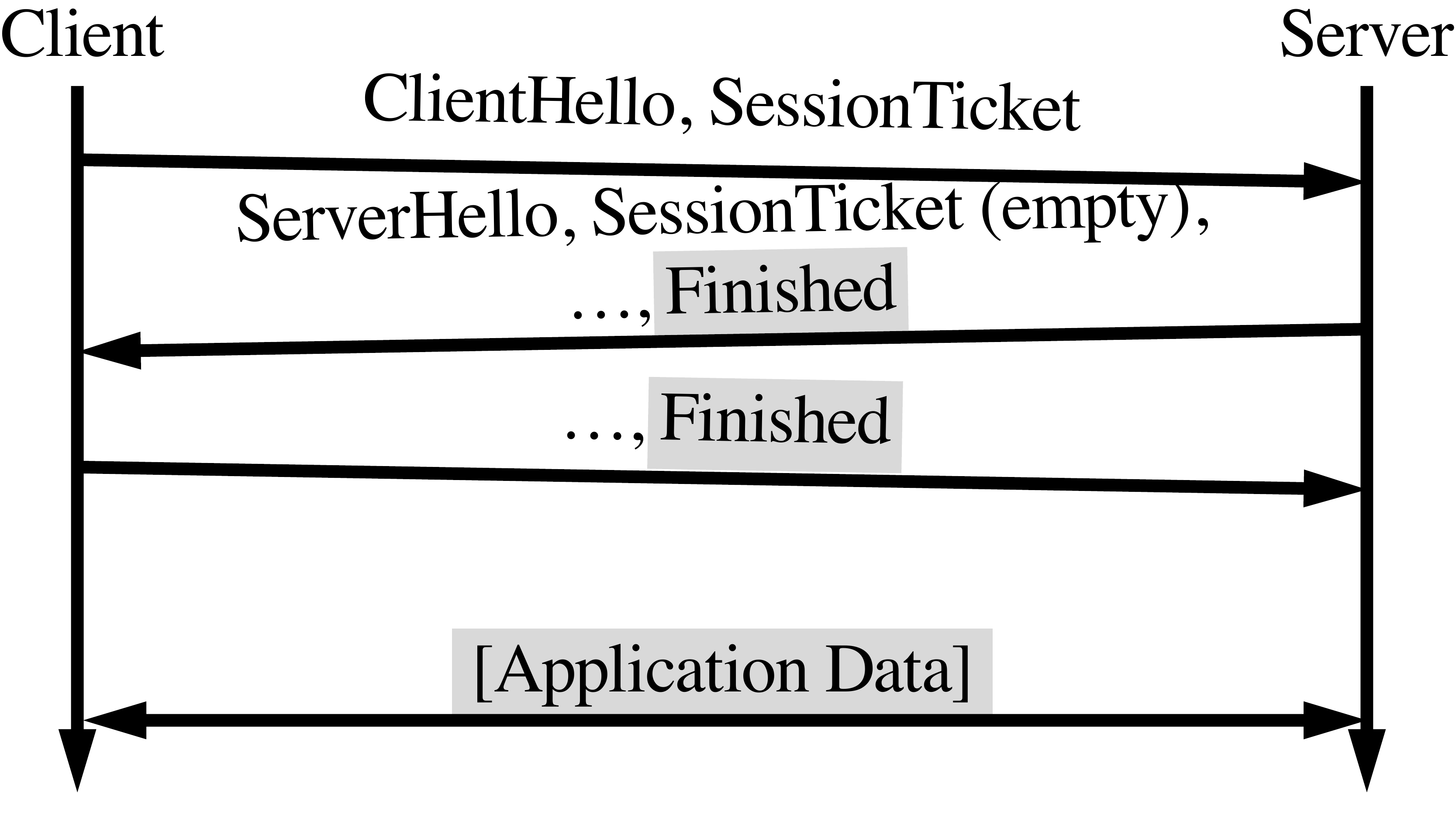}
     \vspace{-10 mm}
     \captionsetup{justification=centering,margin=0.5cm}
     \caption*{c) Session Resumption with\\ Session Ticket}
    \label{fig:tls12_ticket}
\end{minipage}
  \caption{Handshakes in TLS~1.2 with and without session resumption, highlighting \colorbox{mygray}{encrypted} data.}
  \label{fig:tls12_overview}
  \vskip -12pt
\end{figure*}
 
To the best of our knowledge, we are the first to report on the applicability of TLS session resumption for user tracking. The main contributions of our paper are:
\begin{itemize}
  \item We measure session resumption lifetimes of all Alexa Top Million websites and the same for $48$ browsers. We assess the real-world configuration of these mechanisms and derive the possible duration of user tracking, respectively.

\item We introduce the \emph{prolongation attack} that allows to extend the tracking period beyond the session resumption lifetime. 

\item We analyse the impact of the prolongation attack on the resulting tracking periods and on the ratio of permanently trackable users, based on an additional DNS dataset to derive the users' browsing behaviour.
Our results indicate that based on a session resumption lifetime of one day, as standard setting in many popular browsers, the average user can be tracked for up to 8 days. With a session resumption lifetime of seven days, which reflects the recommended upper limit by the draft on TLS version 1.3, even $65$\% of all users in our dataset could be tracked permanently by at least one website in the Alexa dataset.

\item We propose countermeasures that require modifying the TLS standard and the configuration of popular browsers to impede tracking based on TLS session resumption. Most effective is to disable TLS session resumption completely.
\end{itemize}

The remainder of this paper is structured as follows: Section~\ref{background} describes the background on TLS session resumption. Section~\ref{privacy_problems} reviews the privacy problems arising from the usage of TLS and Section~\ref{sec:data_collection} explains the collection of data we used in this paper. Section~\ref{sec:eval} summarizes our major results on tracking via TLS session resumption. Countermeasures are summarized in Section~\ref{sec:countermeasures} and related work is reviewed in Section~\ref{sec:relwork}. Section~\ref{sec:conclusion} concludes the paper.

\section{Background}\label{background}

In this section, we describe session identifiers (IDs) and session tickets as methods to resume a previous TLS connection for TLS up to version~1.2~\cite{TLS12}. Then, we regard session resumption via pre-shared keys (PSK), which is proposed in the draft of TLS~1.3~\cite{TLS13} and not compatible with previous resumption methods. Finally, we compare the presented mechanisms to each other.


\subsection{Session ID Resumption}

In this mechanism, the server assigns a random session ID during the initial handshake with the browser (client). 
Client and server store this session ID along with the session keys and connection states. 
To resume a session, the client sends the stored session ID with the first protocol message (ClientHello) to the server, as shown in Figure~\ref{fig:tls12_overview}\,b). 
If the server recognises the connection and is willing to resume the session, it replies with the same session ID to re-establish the respective session. 

\subsection{Session Ticket Resumption}
This approach is defined in RFC~5077~\cite{ticket} as an extension of the TLS protocol. In its initial ClientHello message, the client is required to express support for session ticket resumption, which will be acknowledged with an appropriate ServerHello response. 
After the key exchange between server and client, the server provides the client with an encrypted session ticket, which is transmitted outside of the TLS encrypted channel. This ticket contains the session keys and connection states, which are encrypted with the private \textit{session ticket encryption key} (STEK) of the server. 
The client stores this session ticket along with the used session key and connection states. 

Upon reconnection, the client includes the session ticket within the ClientHello message as shown in Figure~\ref{fig:tls12_overview}\,c). 
The server then decrypts the session ticket with the STEK and retrieves session key and connection state. If the server accepts the ticket, then the session can be resumed with an abbreviated handshake. 

\subsection{Session Resumption via Pre-Shared Keys}
\begin{figure*}[htpb]
\centering
\begin{minipage}{.33\linewidth}
    \includegraphics[width=\linewidth]{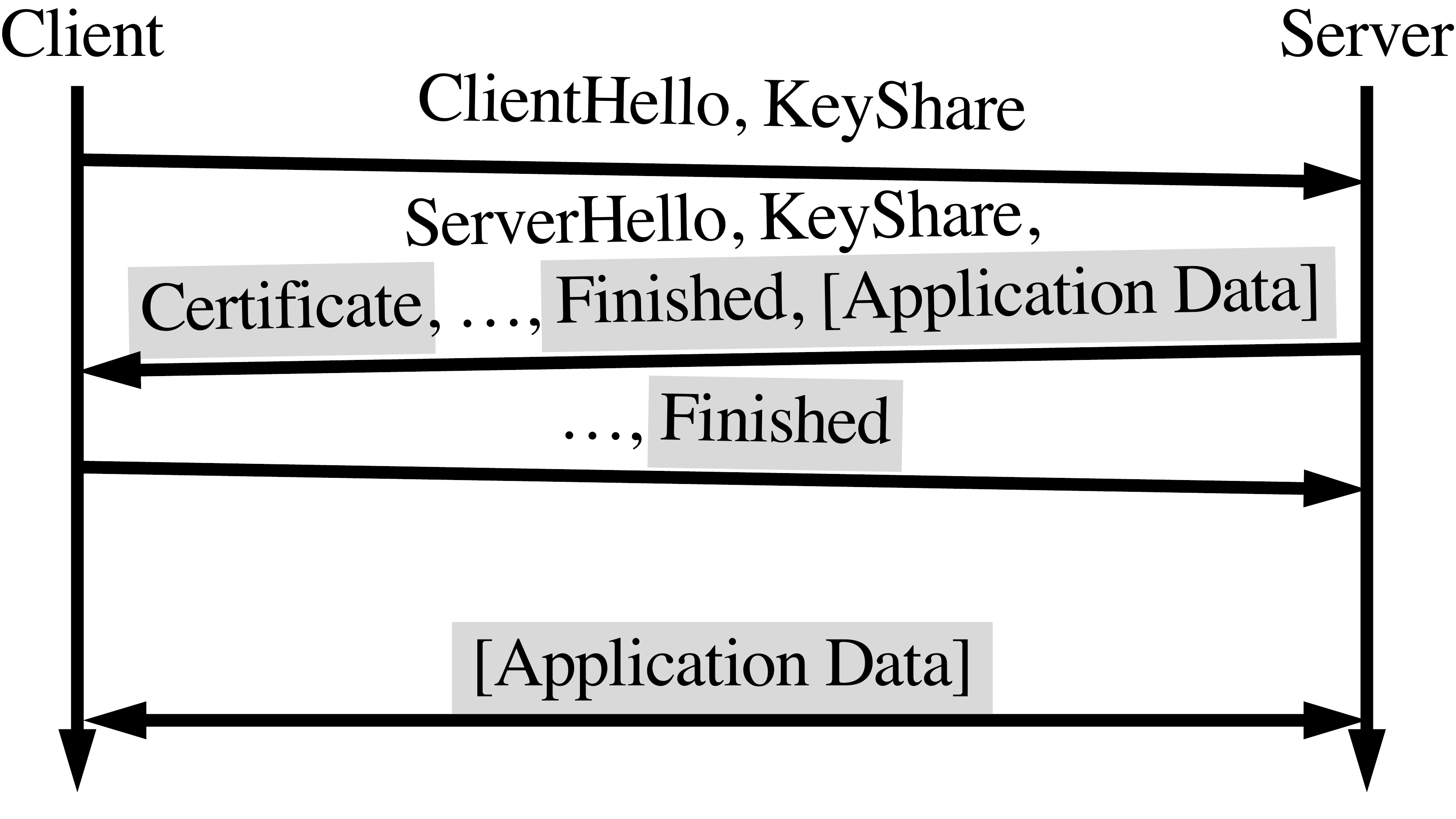}
    \vspace{-10 mm}
    \caption*{a) Full Handshake}
    \label{fig:tls13}
\end{minipage}
\hfill
\begin{minipage}{.33\linewidth}
    \includegraphics[width=\linewidth]{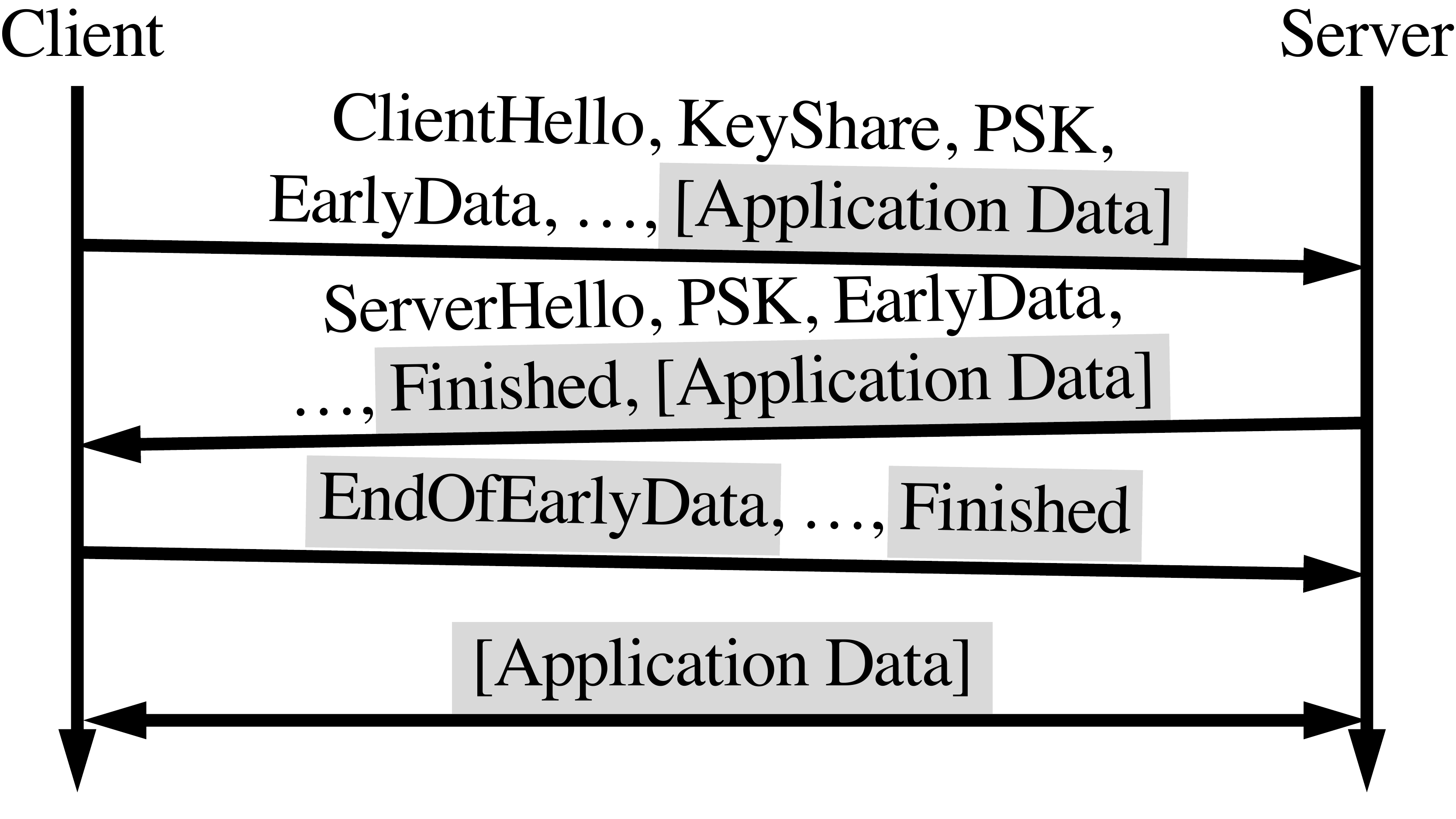}
     \vspace{-10 mm}
    \caption*{b) 0-RTT Session Resumption}
    \label{fig:tls13_0rtt}
\end{minipage}
\hfill
\begin{minipage}{.33\linewidth}
    \includegraphics[width=\linewidth]{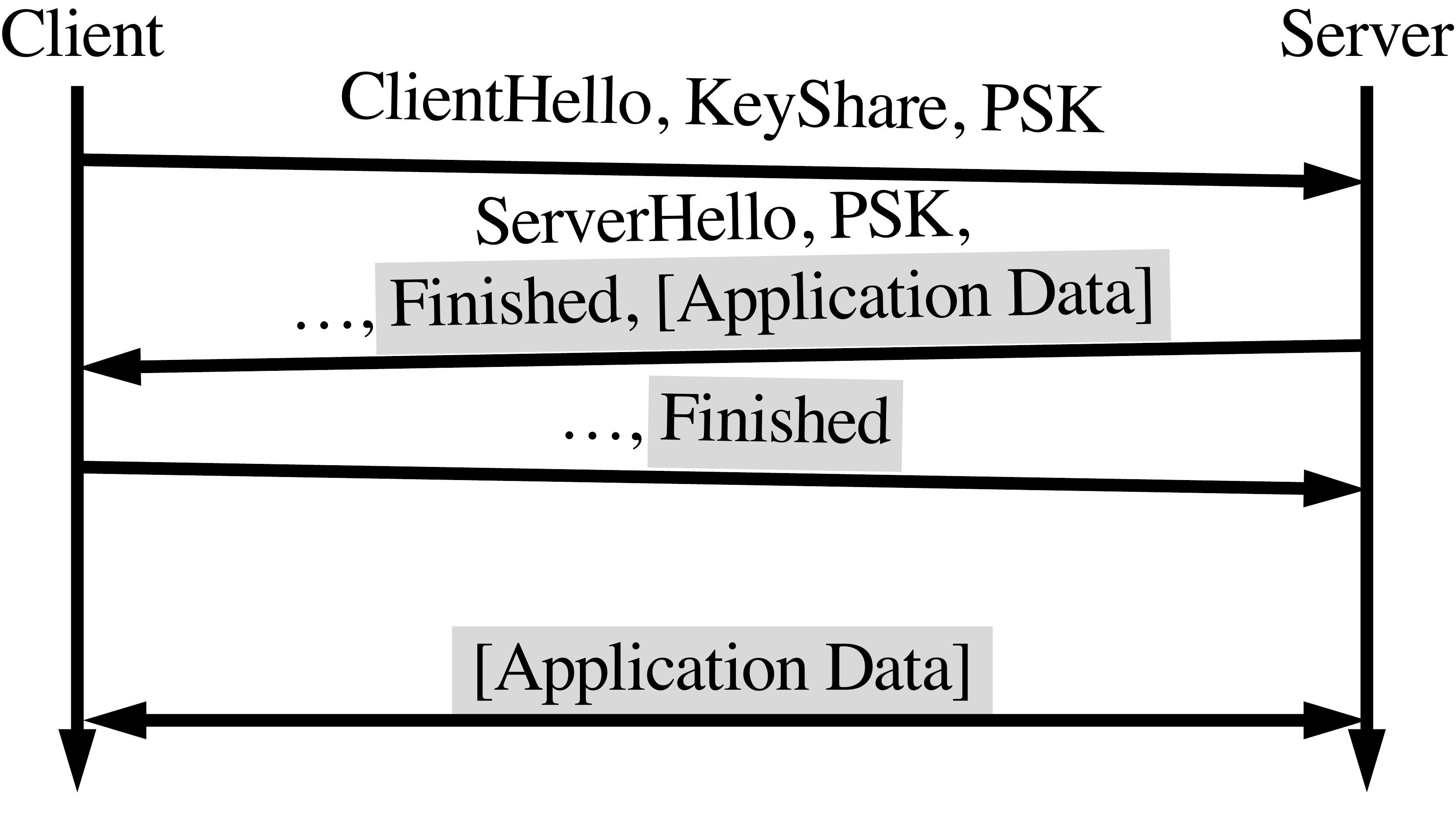}
     \vspace{-10 mm}
    \caption*{c) 1-RTT Session Resumption}
    \label{fig:tls13_1rtt}
\end{minipage}
  \caption{Handshakes in TLS~1.3 with and without session resumption, highlighting \colorbox{mygray}{encrypted} data.}
  \label{fig:tls13_overview}
  \vskip -12pt
\end{figure*}

The draft of TLS~1.3~\cite{TLS13} replaces session IDs and session tickets with the concept of session resumption via pre-shared keys (PSK), which works as shown in Figure~\ref{fig:tls13_overview}. After the initial handshake, the server sends a PSK identity to the client.
The content of the PSK identity depends on the server and may contain a database lookup key or a self-encrypted and self-authenticated ticket.
The client stores this PSK identity along with its own session keys. 

In a subsequent handshake, the client provides this PSK identity within the ClientHello message to the server as seen in Figure~\ref{fig:tls13_overview}\,b) and~\ref{fig:tls13_overview}\,c).
Depending on the content of the PSK identity, the server decrypts the ticket and uses the contained session keys and connection states to resume the session, or the server uses the contained lookup key to find the session keys and connection states in its own database.

\subsection{Comparison of Session Resumption Mechanisms}\label{sec:comparison}

Session resumption via PSK introduces several improvements regarding tracking.
In contrast to session tickets and session IDs, the server sends the PSK identity after an initial handshake through the encrypted TLS channel. Furthermore, the server can issue multiple PSK identities at once, thus each resumption attempt uses a different PSK identity.
These improvements by session resumption via PSK, protect against a correlation of single user sessions by a passive network-based observer.

RFC~5246~\cite{TLS12} recommends a lifetime of session IDs of less than 24 hours, the draft of TLS~1.3~\cite{TLS13} extends this duration to seven days.
No maximum lifetime is specified for session tickets in the RFC~5077~\cite{ticket}. However, servers may provide a hint to the client about their individually supported maximum lifetime.
Thus, for session tickets the lifetime may exceed 24~hours.

The full handshake of TLS~1.3 (see Figure ~\ref{fig:tls13_overview}\,a) requires one round trip less to complete in comparison to TLS~1.2~(see Figure ~\ref{fig:tls12_overview}\,a).
Thus, regarding performance, the methods of session IDs, session tickets, and 1-round-trip time (RTT) session resumption via PSK require the same number of round trips as the full handshake of TLS~1.3, while 0-RTT session resumption via PSK can save one additional round trip.

Another distinction of the discussed resumption mechanisms is their ability to provide forward secrecy.
Forward secrecy describes the property of secure communication protocols, that a compromised long-term cryptographic key does not lead to a compromise of the confidentiality of past sessions. TLS~1.2 and TLS~1.3 generally support forward secrecy by using Diffie-Hellman key establishment to negotiate a temporary, symmetric session key.

From a security perspective, TLS~1.3 improves previous TLS protocol versions by supporting forward secrecy for 1-RTT session resumption (see Figure~\ref{fig:tls13_overview}\,c) through Diffie-Hellman key establishment, however it is not mandatory~\cite{TLS13}. In the case of 0-RTT resumptions, forward secrecy cannot be realised for the first application data transmitted by the client. Thus, session resumption in TLS~1.2 and partially in TLS~1.3 reduces the communication security compared to a fresh TLS session.

Another drawback of 0-RTT resumption is that servers need to implement countermeasures against replay attacks, which TLS itself guards against for other resumption mechanisms. Thus, in TLS~1.3, session resumptions with a reduced number of round trips can only be realised at the cost of reduced security guarantees.

Table~\ref{tab:comparison} provides a brief overview of the differences between the TLS session resumption mechanisms.  

\begin{table*}
  \caption{Comparison of TLS session resumption mechanisms}
  \label{tab:comparison}
  \begin{tabular}{l|cccc}
    \toprule
      &Session ID & Session Tickets & 0-RTT via PSK & 1-RTT via PSK\\
    \midrule
    \textmd{Server stores its own secret TLS state} & yes & no & optional & optional \\
    \textmd{Number of RTT compared to full handshake}& -1 RTT & -1 RTT & -1 RTT & identical\\
    \textmd{Initial handshake contains unencrypted identifier}& yes&yes&no&no\\
    \textmd{Identifier reuse for multiple connections}&yes&yes&should not&should not\\
    \textmd{Forward secrecy}& no&no&no&optional\\
    \textmd{Uses one key for sessions of multiple users}& no&yes&optional&optional\\
     \textmd{Recommended limit of the resumption lifetime}&24h~\cite{TLS12} &>24h~\cite{ticket}&7 days~\cite{TLS13}&7 days ~\cite{TLS13}\\ 
    \bottomrule
  \end{tabular}
\end{table*}

\section{Privacy Problems with TLS Session Resumption}
\label{privacy_problems}

In this section, we describe the impact of session resumption lifetimes on users' privacy. Subsequently, we review the consequences of an unrestricted use of session resumption mechanisms with third-party online trackers.



\subsection{Lifetime of Session Resumption Mechanisms}\label{sec:lifetime}

\textit{Always on} and \textit{always with} are characteristics of mobile devices such as smartphones and tablets that provide a ubiquitous access to the Internet and account for about half of all web browsing activities~\cite{Mobile}. 
A web browser along with its TLS cache can remain active for multiple days in the background of mobile operating systems. Thus, very long session resumption lifetimes of several days or weeks are technically feasible.

\begin{figure}[htpb]
  \centering
  \includegraphics[width=0.4 \textwidth]{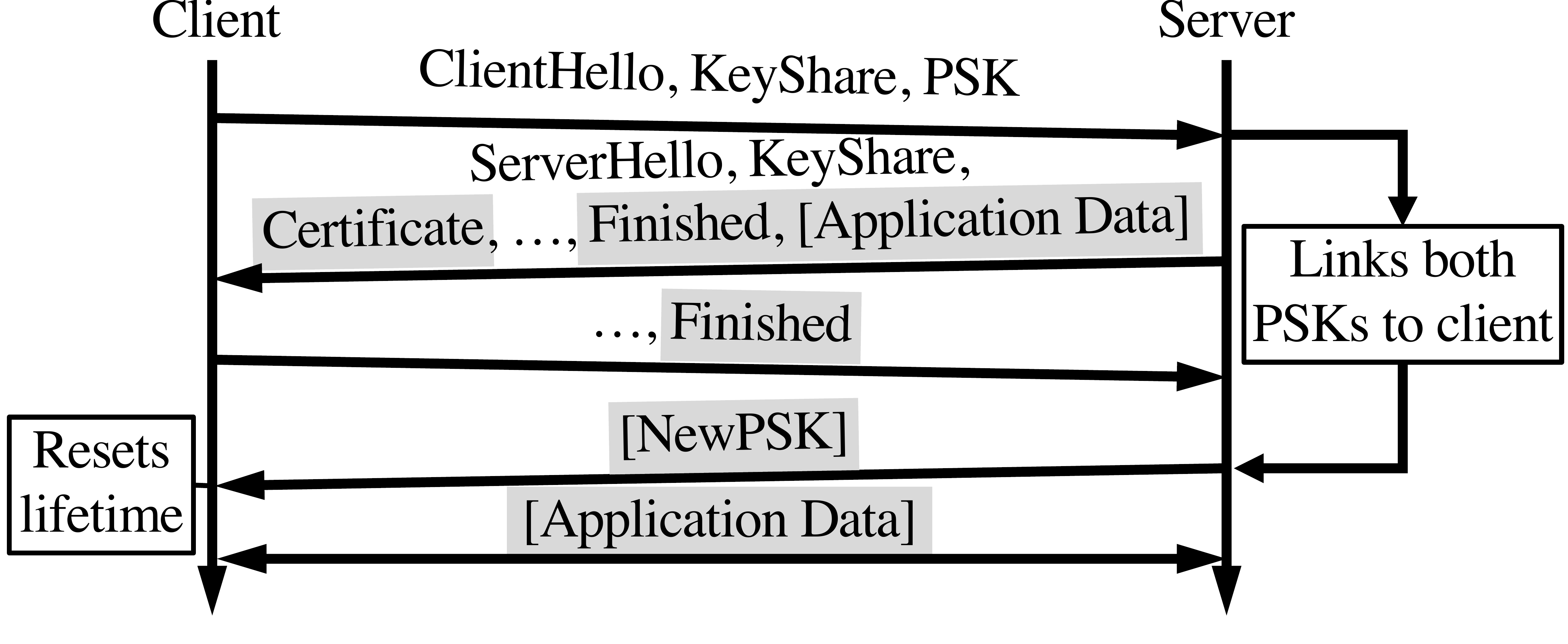}
  \caption{Prolongation attack, where the client attempts a TLS~1.3 1-RTT resumption and the server falls back to a full handshake. The server can link both PSKs to the same user, while the user's resumption lifetime is prolonged with the new PSK.}
  \label{fig:prolongation}
  \vskip -12pt
\end{figure}

Furthermore, the attempt of a client to resume a session by transmitting an identifier to the server leaks the identifier to the server regardless of whether the session is resumed or rejected (see Figure~\ref{fig:prolongation}). Thus, the identifier leakage is sufficient to correlate the initial and the newly established session to the same entity.

To further extend the capability of online tracking, a website might issue a new identifier (session ID, session ticket or PSK identity) on each revisit, and thus track a user indefinitely as long as the time between two visits does not exceed the session resumption lifetime of the user's browser. We refer to this server behaviour as \textit{prolongation attack}.
The TLS standard does not define client behaviour in a way to prevent this attack. While Figure~\ref{fig:prolongation} refers to TLS~1.3 1-RTT, similar attacks apply to 0-RTT and previous TLS versions. 

\subsection{Third-Party Tracking via Session Resumption}\label{sec:third_parties}

Third-party tracking refers to a practice, where a party, other than the targeted website, can track a user's visit. It is a widespread phenomenon on the Internet with an average of 17.7~third-party trackers per website across the Alexa Top 500 categories~\cite{englehardt2016online}. Google with its various hostnames is present on nearly 80\% of the Alexa Top Million Sites~\cite{Alexa, englehardt2016online} and thus can gain deep insights on users' browsing behaviour.

As for tracking via session resumption, third-party trackers can recognise users based on the identifier, which they present to resume a previous TLS session. Thus, the tracker can link multiple observed visits of the user across sites, where the tracker is included as a third-party. However, to distinguish the various first-party sites a user visited, the tracker requires an identifier such as a HTTP referrer or a custom URL per first-party. 

\section{Data Collection}\label{sec:data_collection}

In this section, we describe our various data sources which we use to determine the feasibility of online tracking based on TLS session resumption.
For our empirical analysis, we accumulated the TLS configuration of popular online services and browsers. Furthermore, we investigate web browsing patterns of users based on a DNS traffic data set.

\subsection{Alexa Top Million Data Set}

To get an estimate on the usage of TLS session resumption in the web and to conduct a qualitative analysis of the used session resumption mechanisms we measured the HTTPS behaviour of the Alexa Top Million Sites~\cite{Alexa}. Over a period of 31-days in March and April of 2018, we connected daily to each and every site in the Alexa Top Million on TCP port~443 using the OpenSSL Toolkit~\cite{OpenSSL} and recorded the handshake behaviour.

To probe the configured session resumption lifetime for session IDs and session tickets used by the Alexa Top Hundred Thousand, we revisited each website periodically in intervals of five minutes or less, each time presenting the identical session ID or session ticket, respectively.

To measure which sites share their session ID cache or STEK with other online services, we saved the TLS connection state of a session with one site and tried to pairwise reconnect this state to the other websites in the sample. Subsequently, we created groups of sites that allow mutual session resumption.
This pairwise evaluation causes quadratic cost, therefore we reduced our sample to the Alexa Top Thousand Sites.

We conducted all scans from our university campus and followed best practices of active scanning~\cite{durumeric2013zmap}. 
%

\subsection{Browser Measurements}

To assess the browser behaviour in regard to session resumption we used a sample of 48 browsers for mobile and desktop platforms as shown in Table~\ref{tab:Browser}. This sample includes the most popular browsers with respect to their global market share~\cite{statcounter, browser_stat, netmarketshare}, which were publicly accessible for either iOS, Android, and/or desktop operating systems. Besides the most popular browsers, we included Tor Browser, Orbot, Brave, Cliqz, JonDoBrowser, and Ghostery Privacy Browser as explicitly privacy-friendly browsers to our sample.

To gather the configured maximum resumption lifetimes of each browser and for each different resumption mechanisms, we used a test website with a custom JavaScript probe. We attempted to resume a session after varying intervals of up to 24 hours since the initial handshake. On each connection attempt, the server checks and records if the initially established session resumption identifier is transmitted.
We tried only one session resumption per initial handshake to avoid potential side effects like a prolongation of the browser's resumption lifetime.

\begin{figure}[htpb]
  \centering
  \includegraphics[width=0.4 \textwidth]{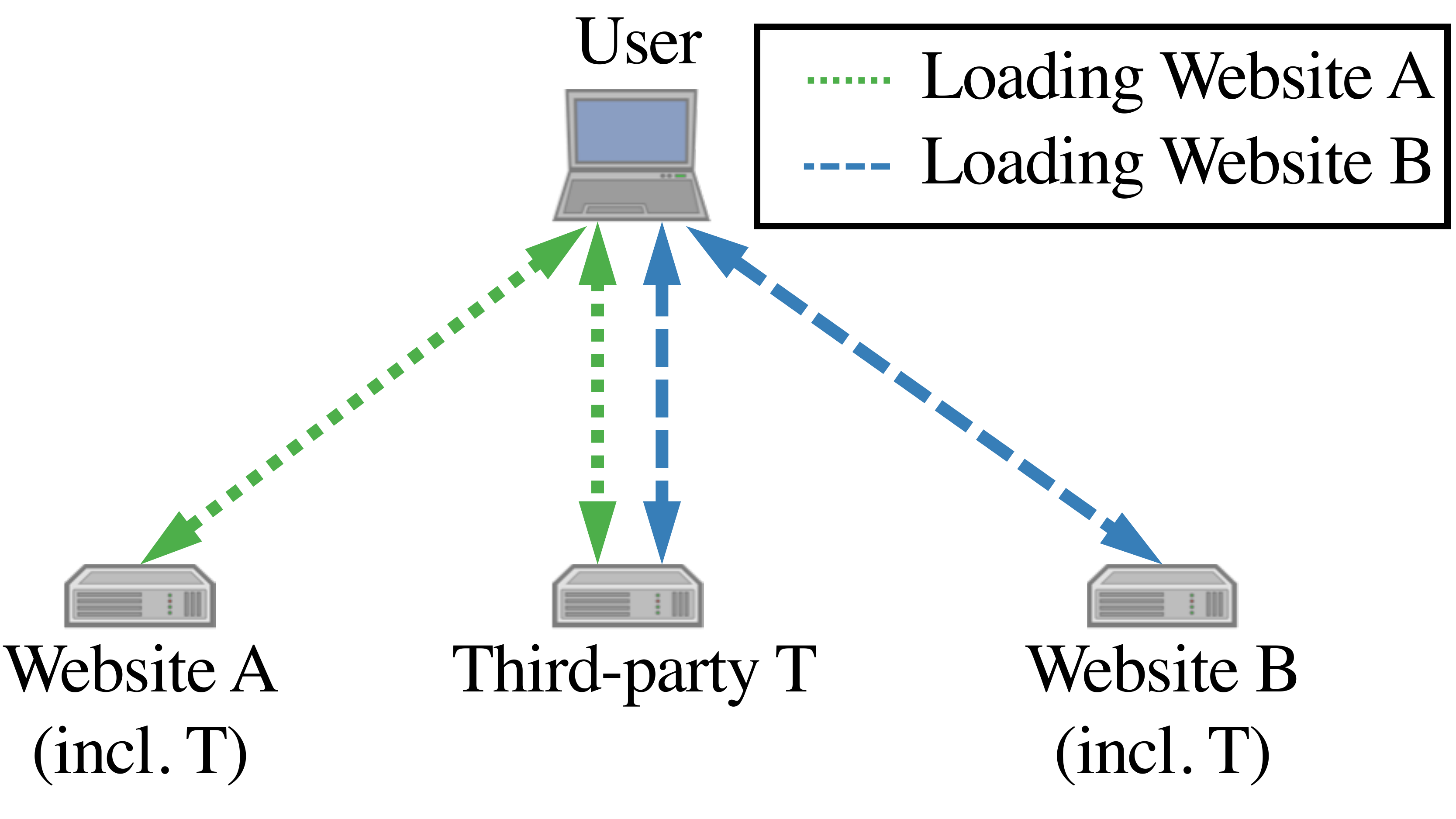}
  \caption{Testbed to measure browser behaviour in regard of third-party tracking.}
  \label{fig:third_party}
  \vskip -12pt
\end{figure}

To test if browsers enable third-parties to link user activities on different websites as described in Section~\ref{sec:third_parties}, we used a testbed as illustrated in Figure~\ref{fig:third_party}. The browser consecutively retrieves the websites A and B, which include the same third-party T. We observe, whether a session resumption with T is possible from the context of different first parties A and B.

\subsection{DNS Data Set}
\label{sec:dns-data-set}

To estimate the impact of the presented prolongation attack (see Section \ref{sec:lifetime}) on the tracking period of a user, we evaluate real-world browsing patterns. Furthermore, these browsing patterns allow us to approximate the ratio of resumed revisits for a given session resumption lifetime.

We obtained the data set used in~\cite{herrmann2013behavior, kirchler2016tracked, herrmann2016behavior}, which contains the pseudonymized DNS traffic logs of 3862 students over a period of two months between April 30, 2010 and June 29, 2010. The data set originates from the student housing network at the University of Regensburg.

Our sample of DNS logs contains only the fraction of Internet traffic that originates from the students' unique and static IP address accessible within their room. Note that this sample might not cover the whole traffic of all users and thus our conclusions might only describe a lower boundary in regard to user tracking via TLS session resumption.
A descriptive statistic of the used DNS data set can be found in~\cite{herrmann2013behavior}.

We restricted our evaluations to a pseudonymized DNS data set to address ethical concerns. Besides a pseudonym for the source IP address, we also pseudonymized target hostnames on a per user basis. This approach prevents background knowledge attacks that use knowledge about the browsing history of another user within this data set.

Consequently, each record in our data set consists of a pseudonym for the source IP address, the query time, a pseudonym for the target hostname, and query type. For our analysis we considered only DNS queries of regular name resolution for IPv4 addresses as well as IPv6 addresses.

\section{Evaluation}\label{sec:eval}

To evaluate the feasibility of tracking based on TLS session resumption, we analyse in this section the configuration of servers and browsers and to which extent they restrict the session resumption mechanisms.
Afterwards, we investigate real-world browsing patterns and check whether the technical restrictions of browsers are suitable to protect users' privacy against online tracking services.  

\subsection{Evaluation of Server Configurations}\label{sec:server_config}

The feasibility of TLS session resumption as a tracking mechanism depends considerably on the configuration of the server. In this section, we investigate the adaptation of TLS session resumption mechanisms, real-world configurations for the session resumption lifetime and security-related configuration issues.

\subsubsection{Adoption of TLS Session Resumption Mechanisms}

\begin{figure}
\centering
\vskip -12pt
\includegraphics[trim={0 1.5cm 0 0cm},clip,width=0.4 \textwidth]{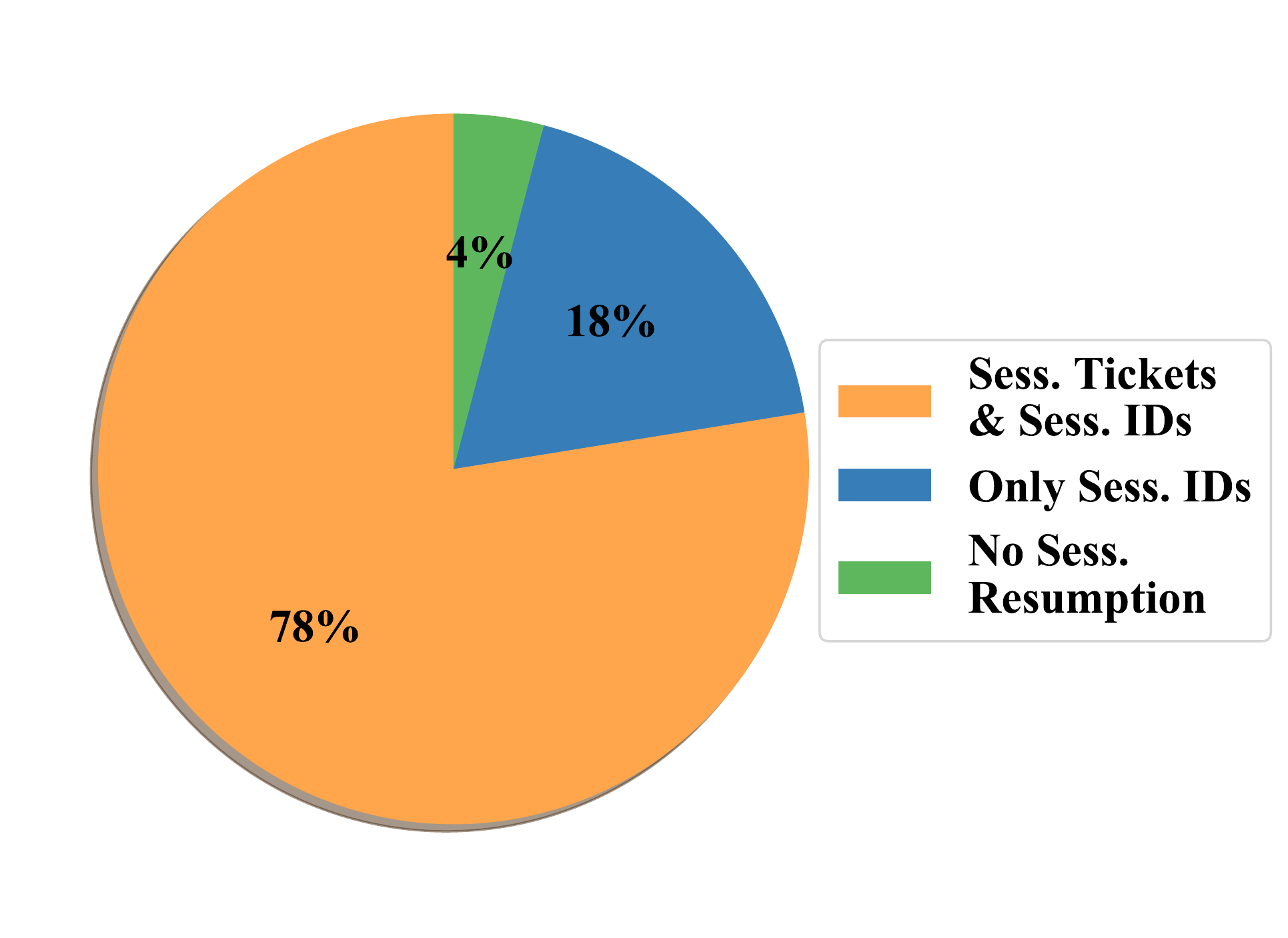}
\vskip -12pt
\caption{Supported session resumption mechanisms for TLS-enabled sites in Alexa Top Million on 15th April 2018.} 
\label{fig:overview_resumption}
\vskip -12pt
\end{figure}

Figure~\ref{fig:overview_resumption} shows the support of TLS-enabled Alexa Top Million Sites for session resumption based on IDs or tickets. This measurement from the 15th April 2018 does not include session resumption via PSK because TLS~1.3 was still a draft at that point.
In total, we found 691\,280 sites among the Alexa Top Million that supported TLS.
$95.9\%$ of these sites do either support session IDs or both IDs and tickets.
The remaining 28\,236 sites do not support TLS session resumption by, for example, providing an empty string as an ID which does not allow to resume a session.

With 536\,088 websites, $77.6\%$ of all TLS-enabled Alexa Top Million Sites do also support session tickets. Note that, if a client-server pair supports both session resumption mechanisms, then TLS session tickets will be the preferred resumption mechanism~\cite{ticket}. 

\subsubsection{Lifetime of Session Resumption Mechanisms}

The server can include a \textit{lifetime hint} along with the session ticket or PSK identity. If the browser respects lifetime hints, it will only try to resume previous sessions within this hinted lifetime.

\begin{figure}
\centering
\includegraphics[width=0.45 \textwidth]{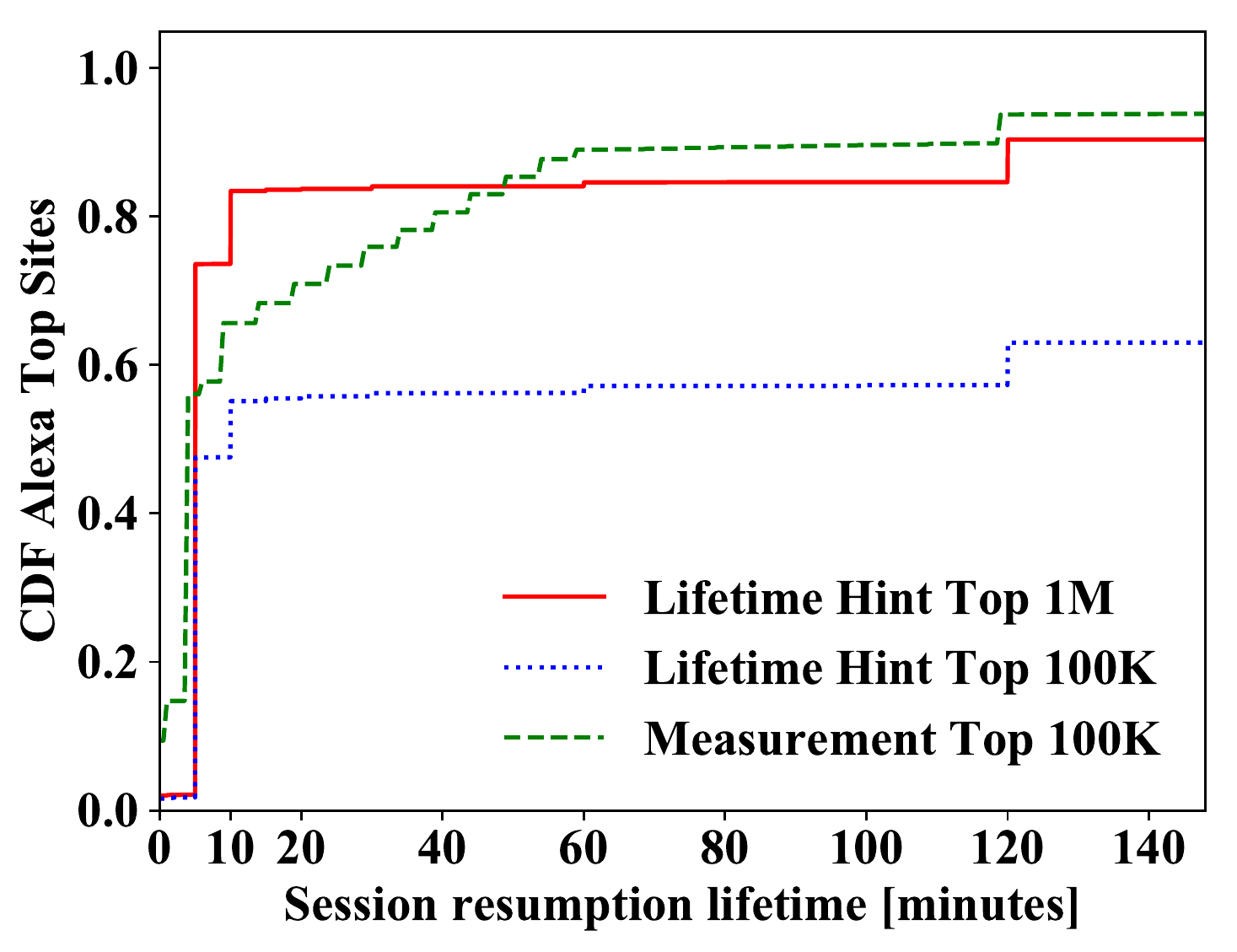}
\caption{Cumulative distribution of Alexa Top Sites over short hinted and measured lifetimes of session tickets.}
\label{fig:CDF_ticket}
\vskip -12pt
\end{figure}

The lifetime hints of TLS session tickets for the Alexa Top Million set (solid, red line) and Alexa Top Hundred Thousand set (dotted, blue line) are shown as cumulative distribution in Figure~\ref{fig:CDF_ticket}. This plot is normalised to the total number of obtained tickets, which were $535\,306$ and $56\,407$ for the respective sets on the 24th March 2018. With $46\%$ and $71\%$, a lifetime hint of five minutes is the most popular configuration within the Alexa Top Hundred Thousand and Alexa Top Million respectively.

In both sets, around $2\%$ of all sites use a lifetime hint of zero seconds. We interpret this configuration as erroneous because this effectively prevents session resumption. Instead, server operators should deactivate the session ticket extension to save resources for both server and client if they choose not to support session tickets.

\begin{figure}
\centering
\includegraphics[width=0.45 \textwidth]{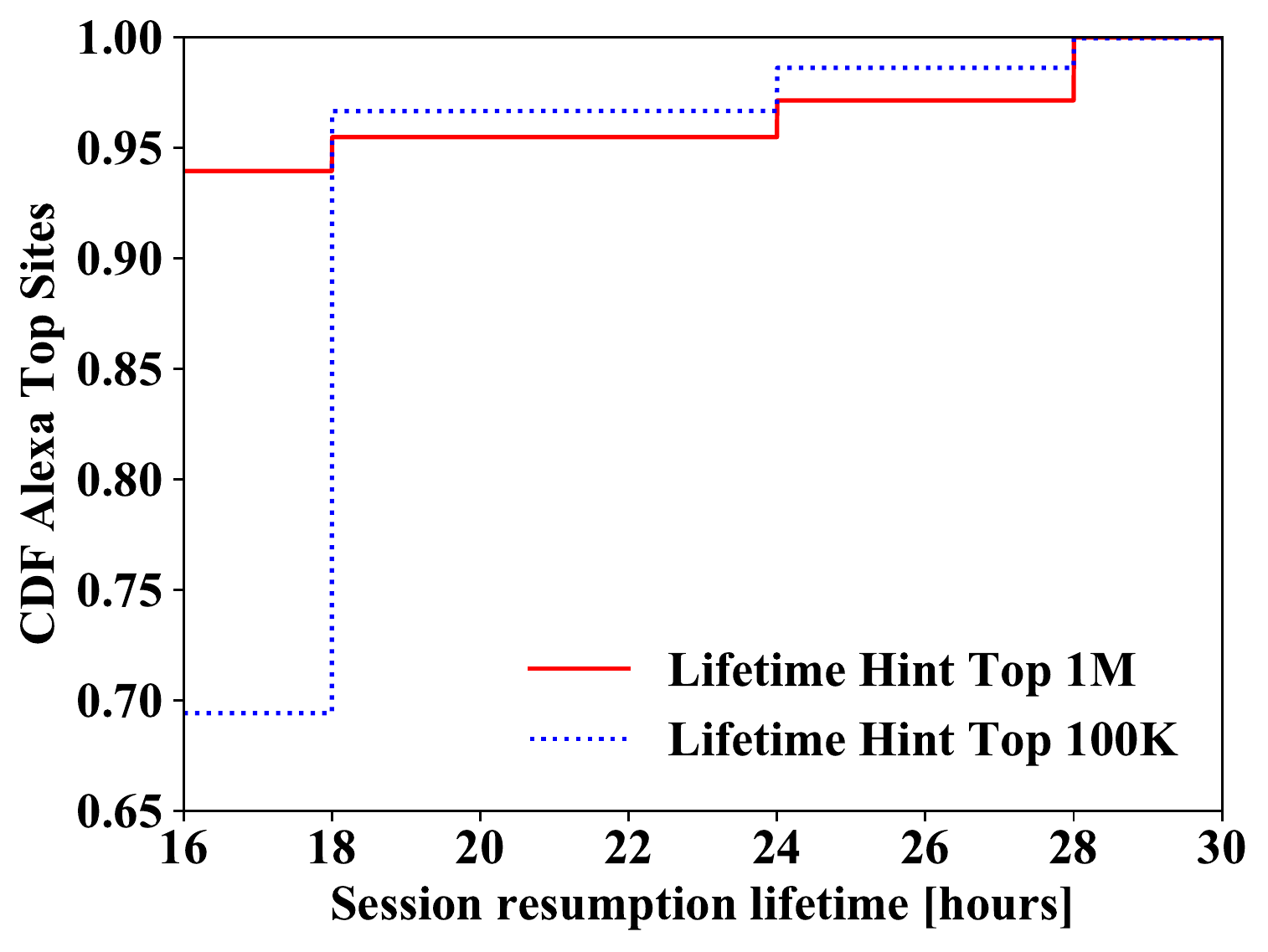}
\caption{Cumulative distribution of Alexa Top Sites over long hinted lifetimes of session resumption tickets.}
\label{fig:CDF_ticket_zoom}
\vskip -12pt
\end{figure}

We observe, that more than $80\%$ of TLS session ticket enabled sites within the Alexa Top Million chose lifetime hints of less than or equal to ten minutes. However, around $10\%$ of the remaining sites use lifetime hints larger than 24~hours. Google and Facebook as market leaders in behavioural advertising show particularly large hinted session resumption lifetimes. Facebook's lifetime hint of 48~hours is above the $99.99\%$ percentile of all session ticket hints that we collected in our scans. Google and various of its domains are configured to a ticket lifetime of 28~hours, which is above the $97,13\%$ percentile in the Alexa Top Million Sites~(see Figure~\ref{fig:CDF_ticket_zoom}).

\begin{figure}
\centering
\includegraphics[width=0.45 \textwidth]{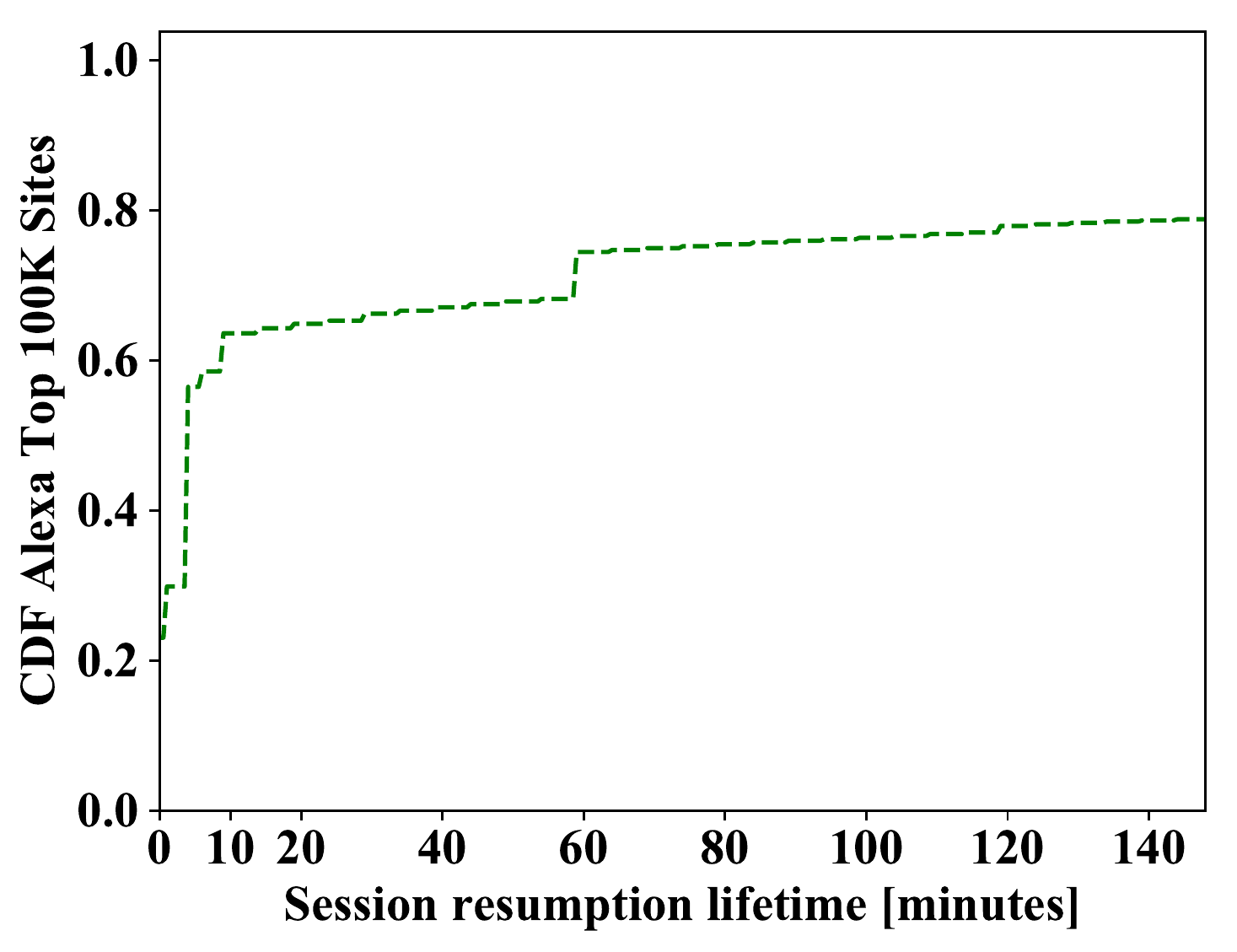}
\caption{Cumulative distribution over the measured session resumption lifetime with session IDs.}
\label{fig:CDF_id}
\vskip -12pt
\end{figure}

To validate these results about lifetime hints, we measured the maximum session resumption lifetimes for session IDs and session tickets.
We limited our sample size to the Alexa Top Hundred Thousand Sites for practicality reasons.
Figure~\ref{fig:CDF_ticket} and \ref{fig:CDF_id} show these results as green, dashed line for tickets and IDs, respectively. We find, that around $10\%$  of the sites do not allow a resumption with session tickets immediately after the initial visit, while, for session IDs these are $23\%$ of the websites. We observe in both figures, that approximately $40\%$ of the sites support maximum resumption lifetimes above five minutes. More than $20\%$ of websites within the Alexa Top Hundred Thousand that support session IDs allow resumptions after more than two hours.

As can be seen in Figure~\ref{fig:CDF_ticket}, our measurement of the maximum resumption lifetime deviates within a range of 10 to 55~minutes from the blue, dotted plot of the corresponding lifetime hints. So far, we are not able to explain this zig-zag shape of the green plot. 

\subsubsection{Security Issues of TLS Server Configurations}\label{sec:server_security}

In this section, we empirically measure TLS state sharing, where multiple sites share their cryptographic secrets for user sessions. Furthermore, we evaluate the lifetime of \textit{Session Ticket Encryption Keys} (STEK). These two measurements allow us to assess the vulnerable period and the number of vulnerable websites in case of a compromise of a site's STEK.


\begin{figure}
\centering
\includegraphics[width=0.45 \textwidth]{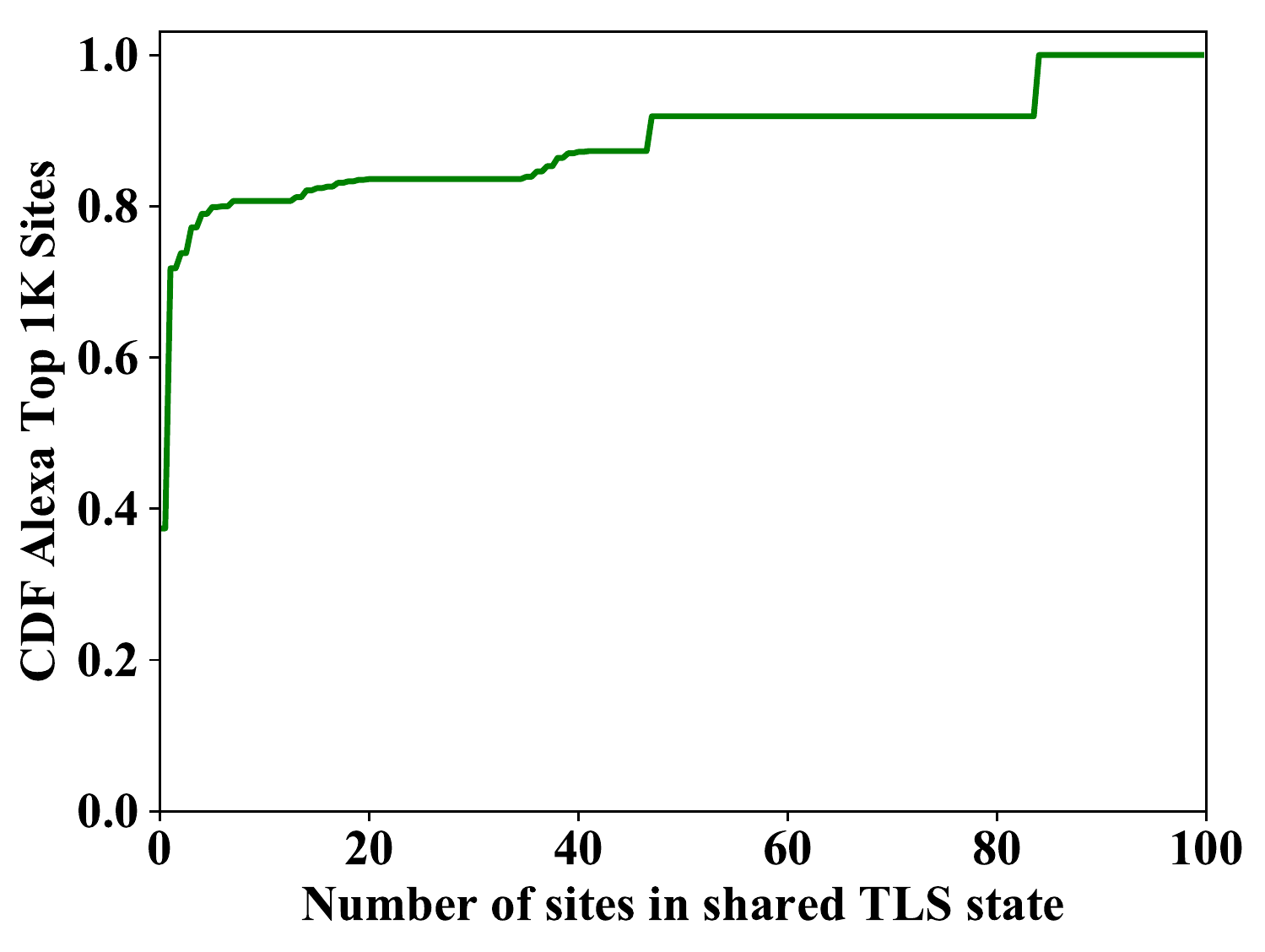}
\caption{Cumulative distribution over the size of shared TLS states among sites within the Alexa Top Thousand.}
\label{fig:sharedTLS}
\vskip -12pt
\end{figure}

A passive, network-based observer in possession of a compromised STEK can decrypt the initial and the resumed sessions in TLS 1.2 due to the missing forward secrecy for session resumption~(see Section~\ref{sec:comparison}).
This issue becomes more severe as STEKs are shared across multiple hostnames, which increases the number of potentially affected sessions as well as the attack surface.

We measured the shared TLS state by attempting to resume sessions of website A, where A is within Alexa Top Thousand, on every other site of the Alexa Top Thousand. 

Figure~\ref{fig:sharedTLS} shows that $72\%$ of the Alexa Top Thousand do either not support session resumption or do not share their TLS state with any other site of the Alexa Top Thousand. However, around $16\%$ of websites share their TLS state with at least 25 other sites of the Alexa Top Thousand. The largest shared TLS state within the Alexa Top Thousand counts 84~sites and belongs to Google.


RFC 5077~\cite{ticket} recommends a scheme for the construction of TLS session tickets, which includes an identifier for the respective STEK. With daily scans of the Alex Top Million Sites for 31-days, we obtained a sample of over 16~million session tickets. In the next step, we assigned each site their corresponding set of STEK identifiers. 

\begin{figure}
\centering
\includegraphics[trim={0 1.5cm 0 1.5cm},clip,width=0.4 \textwidth]{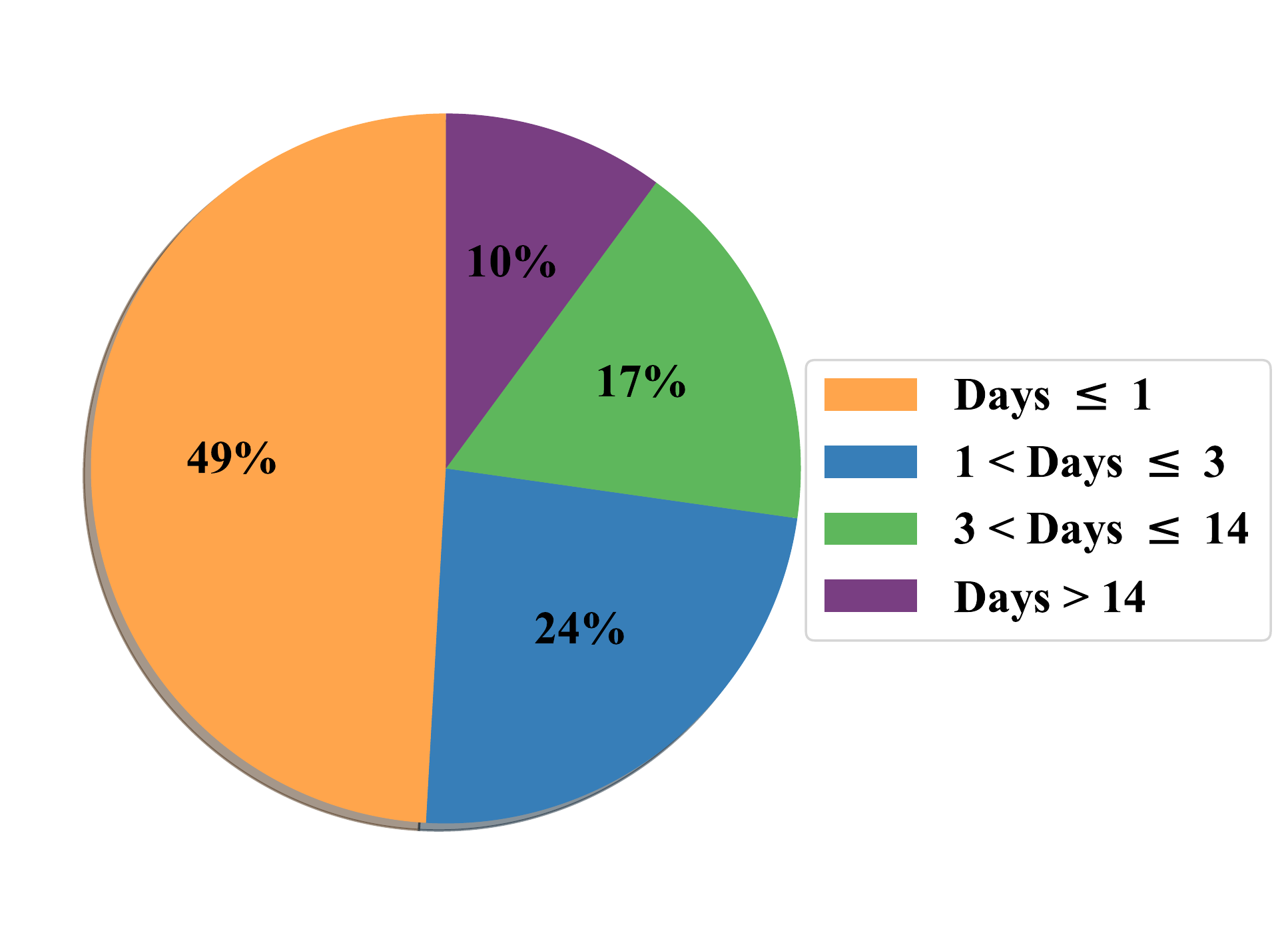}
\vskip -12pt
\caption{Distribution of Alexa Top Million Sites over the period for which they encrypt session tickets with the identical cryptographic key.}
\label{fig:Stek_lifetime}
\vskip -12pt
\end{figure}

Figure~\ref{fig:Stek_lifetime} shows, that $73\%$ of the TLS session ticket enabled Alexa Top Million Sites to change their STEK within three days. However, $10\%$ of the websites with support for session tickets did not change their STEK within two weeks.
Consequently, a compromised STEK from such a website allows a network-based observer to decrypt all sessions that use session tickets with this site for at least two weeks.

Note that TLS implementations might deviate from the recommended construction scheme for session tickets~\cite{ticket}. Our setup expects session tickets to follow the recommended structure, deviating websites are therefore falsely classified as exchanging STEKs with every new ticket. Thus, Figure~\ref{fig:Stek_lifetime} shows only a lower boundary of websites that change their STEK less frequently than every 24~hours.

\subsection{Evaluation of Browser Configurations}\label{sec:browser_config}

In this section, we investigate the default configuration of popular web browsers to determine the feasibility of user tracking via TLS session resumption.
First, we report on the session resumption lifetime of browsers for session IDs and tickets.
Then, we evaluate the capability of third-parties to track users across different first-party websites.

 \begin{table}[htbp]
   \caption{TLS session resumption configuration of browsers}
  \label{tab:Browser}
 \centering
 \begin{tabular}{|c|l|r|r|r|}
 \hline
\rotatebox[origin=c]{90}{Plt.} & \multicolumn{1}{c|}{Browser} & \multicolumn{1}{c|}{\thead{Lifetime for\\Sess. ID}} & \multicolumn{1}{c|}{\thead{Lifetime for\\Sess. Ticket}} & \multicolumn{1}{c|}{\thead{Third-party\\tracking}}\\
 \hline
   \cellcolor[gray]{0.8}&360 Sec. Bro. v9.1&600 min&540 min&blocked\\
   \cellcolor[gray]{0.8}&Amigo v61.0&30 min&30 min&viable\\
   \cellcolor[gray]{0.8}&Brave&30 min&30 min&viable\\   
   \cellcolor[gray]{0.8}&Chrome v66.0&60 min&30 min&viable\\
    \cellcolor[gray]{0.8} &Cliqz&1 day&10 min&viable\\ 
   \cellcolor[gray]{0.8}&Coc Coc v68.4&30 min&30 min&viable\\
   \cellcolor[gray]{0.8} &Comodo Dra. v63&30 min&30 min&viable\\ 
   \cellcolor[gray]{0.8}&Firefox v59.0&1 day&1 day&viable\\
   \cellcolor[gray]{0.8}&Internet Expl. v11&600 min&600 min&viable\\
  \cellcolor[gray]{0.8} &JonDoBrowser& - & - &blocked\\ 
     \cellcolor[gray]{0.8}&Konqueror v5.0&30 min&30 min&blocked\\
  \cellcolor[gray]{0.8}&K-Meleon v75.1&1 day&1 day&viable\\
     \cellcolor[gray]{0.8}&Lunascape v6.15&600 min&540 min&viable\\
        \cellcolor[gray]{0.8}&Maxthon v5.2&30 min&30 min&viable\\
           \cellcolor[gray]{0.8}&Microsoft Edge v41&600 min&600 min&blocked\\
    \cellcolor[gray]{0.8}&Opera v52&30 min&30 min&viable\\
   \cellcolor[gray]{0.8}&Pale Moon v27&1 day&10 min&viable\\
      \cellcolor[gray]{0.8}&QQ Browser v10&30 min&30 min&viable\\
    \cellcolor[gray]{0.8}&QupZilla v2.2.6&30 min&30 min&viable\\
    \cellcolor[gray]{0.8}&Safari v11.1&1 day&1 day&viable\\
       \cellcolor[gray]{0.8}&SeaMonkey v2.49&1 day&1 day&viable\\
    \cellcolor[gray]{0.8}&Sleipnir v6.2&30 min&30 min&blocked\\
 \cellcolor[gray]{0.8}&Sogou Expl. v8&30 min&30 min&viable\\
 \cellcolor[gray]{0.8}&SRWare Iron v65&30 min&30 min&viable\\
    \cellcolor[gray]{0.8} &Tor Browser&-&-&blocked\\ 
 \cellcolor[gray]{0.8}&UC Browser v7.0&30 min&30 min&viable\\
 \parbox[t]{2mm}{\multirow{-27}{*}{\rotatebox[origin=c]{90}{\cellcolor[gray]{0.8}Desktop}}} &Vivaldi v1.14&30 min&30 min&viable\\
  \cellcolor[gray]{0.9}&Amigo v1.10.187&60 min&30 min&viable\\
 \cellcolor[gray]{0.9}&Android Bro. v7.1.2&30 min&30 min&viable\\ 
  \cellcolor[gray]{0.9}&Brave v1.0.42&60 min&30 min&viable\\ 
   \cellcolor[gray]{0.9}&Cheetah Bro. v5.22&60 min&30 min&viable\\
 \cellcolor[gray]{0.9}&Chrome v61.0&30 min&60 min&viable\\
  \cellcolor[gray]{0.9}&Cliqz v1.6.2&60 min&30 min&viable\\
 \cellcolor[gray]{0.9}&Firefox v56.0&1 day&20 min&viable\\ 
    \cellcolor[gray]{0.9}&Ghostery Priv. v1.3&30 min&30 min&viable\\ 
 \cellcolor[gray]{0.9}&Maxthon v4.5.10&30 min&30 min&viable\\ 
 \cellcolor[gray]{0.9}&Opera Mini v30.0&15 sec&15 sec&viable\\ 
  \cellcolor[gray]{0.9}&Orbot v16.0.0&-&-&blocked\\  
 \cellcolor[gray]{0.9}&QQ Bro. v1.2.0&1 day&1 day&viable\\ 
 \cellcolor[gray]{0.9}&Samsung Intern. v6&50 min&50 min&viable\\ 
 \cellcolor[gray]{0.9}&Sleipnir v 3.5.7&60 min&30 min&viable\\ 
 \cellcolor[gray]{0.9}&SRWare Iron v61.0&60 min&30 min&viable\\ 
  \cellcolor[gray]{0.9}&UC Browser v12.0&1 day&30 min&viable\\ 
 \parbox[t]{2mm}{\multirow{-17}{*}{\rotatebox[origin=c]{90}{ \cellcolor[gray]{0.9}Android}}} &Yandex Bro. v18.1&50 min&30 min&viable\\
    \cellcolor[gray]{0.8}&Chrome v62.0&120 min&-&viable\\
 \cellcolor[gray]{0.8} &Firefox v9.2&120 min&-&viable\\ 
  \cellcolor[gray]{0.8} &Opera v16.0&120 min&-&viable\\ 
 \parbox[t]{2mm}{\multirow{-4}{*}{\rotatebox[origin=c]{90}{\cellcolor[gray]{0.8}iOS 11}}} &Safari v11.0&120 min&-&viable\\
 \hline
 \end{tabular}
 \end{table}

\subsubsection{Lifetime of Session Resumptions Mechanisms}\label{sec:browser_lifetime}

The results of measuring the session resumption lifetime are displayed in Table~\ref{tab:Browser}. We note that only three browsers from the privacy-friendly group do not support TLS session resumption, the remaining 45 brows\-ers all support at least one resumption mechanisms. 

We observe, that two-thirds of all tested browsers allow only for a session resumption lifetime of up to 60~minutes for both resumption mechanisms. Our measurement period is limited to 24~hours, therefore a stated one-day period might, in fact, be longer. 

Some browsers such as Chrome, Opera, Firefox, and Safari were tested on multiple platforms. For these browsers, we find that the configuration of the session resumption lifetime is not consistent across the investigated platforms. For example, the desktop version of Safari is configured to a session resumption lifetime of 24~hours, while the iOS version only supports resumptions for session IDs of up to 120~minutes.

Furthermore, Table~\ref{tab:Browser} shows that the lifetime for both investigated resumption mechanisms differ for Cliqz (desktop), Pale Moon (desktop), Firefox (Android), and UC Browser (Android) by more than 23~hours. For browsers on iOS, we observe a homogeneous behaviour in our measurements, which can be explained by Apple's requirement that all browser apps in the App Store use Apple's WebKit framework as rendering engine~\cite{Webkit}.  

\subsubsection{Third-party Tracking}

We analysed whether the default configuration of browsers allows resuming TLS sessions in the context of different first-party websites. Table~\ref{tab:Browser} shows, that only the desktop browsers 360 Secure Browser, Konqueror, Microsoft Edge, and Sleipnir restrict the session resumption support for third-parties. Note, that the privacy-friendly browsers JonDoBrowser, Tor Browser, and Orbot do not support TLS session resumption mechanisms and thus also prevent third-party tracking in this regard.

Our results show, that third-party tracking via TLS session resumption is feasible for the large majority of investigated popular browsers. However, our results about the session resumption lifetime~(see Section~\ref{sec:browser_lifetime}) indicate the session resumption lifetime is limited within the majority of investigated browsers. An explanation of our results regarding third-party tracking could be that browser vendors are mostly unaware of third-party tracking via TLS session resumption.

\subsection{Evaluation of Real-World User Traffic}


In this section, we use real-world data from a DNS traffic data set to assess the impact of different session resumption lifetimes on the achievable length of tracking periods as well as the share of permanently trackable users, respectively using the prolongation attack.
Subsequently, we analyse the impact of a chosen session resumption lifetime on the expectable frequency of session resumptions for revisits. 

\subsubsection{Assumptions and Limitations}
Lacking access to a comprehensive data set of actual web browsing activity, we resort to a DNS traffic data set as described in Section~\ref{sec:dns-data-set}.
For our evaluation, we assume that all DNS name resolution queries in the data set led to a TLS session.
As discussed in Section~\ref{privacy_problems}, we reason that \textit{always on} devices such as smartphones or tablets can achieve runtime durations for browsers and their TLS cache of several days for average users.
Therefore, we assume that users would not have cleared the TLS cache of their browser within the tracking period.
Note, that the following evaluations only approximate actual website visits since we cannot account for DNS caching effects.

\subsubsection{Longest Consecutive Tracking Period}

Tracking mechanisms become more capable the longer it is possible to link user behaviour to a known entity.
Recall that by using the prolongation attack as described in Section~\ref{sec:lifetime}, the period in which user activities can be linked via session resumption identifiers can be extended beyond a single session resumption lifetime.
The extent to which this prolongation is possible depends on the time between consecutive visits of a website.

\begin{definition}
  In the context of session resumption we use the term \textit{consecutive tracking period} to refer to a sequence of visits \(v_1, \dots, v_n\) to an online service where no interval between two visits exceeds the given session resumption lifetime \(l\). More formally, we define the predicate \(P\) as
  \[P(v_1, \dots, v_n, l) \Leftrightarrow |t_i - t_{i+1}| \le l, \forall i \in \{1, \dots, n-1\},\]
  where \(t_i\) denotes the time of visit \(v_i\).
\end{definition}

\begin{definition}
  We further define the \textit{longest consecutive tracking period} \(lctp\) of a sequence of visits \(v_1, \dots, v_n\) given a session resumption lifetime \(l\) as the length of the longest subsequence of visits that still fulfils the consecutive tracking period property, or more formally as
  \[ lctp(v, l) \mapsto \max_{i,j \in \{1, \dots, n\}, i < j} |t_i - t_j| [P(v_i,\dots,v_j,l)],\]
  where \(t_i\) again denotes the time of visit \(v_i\).
\end{definition}

\begin{figure}
\centering
\includegraphics[width=0.45 \textwidth]{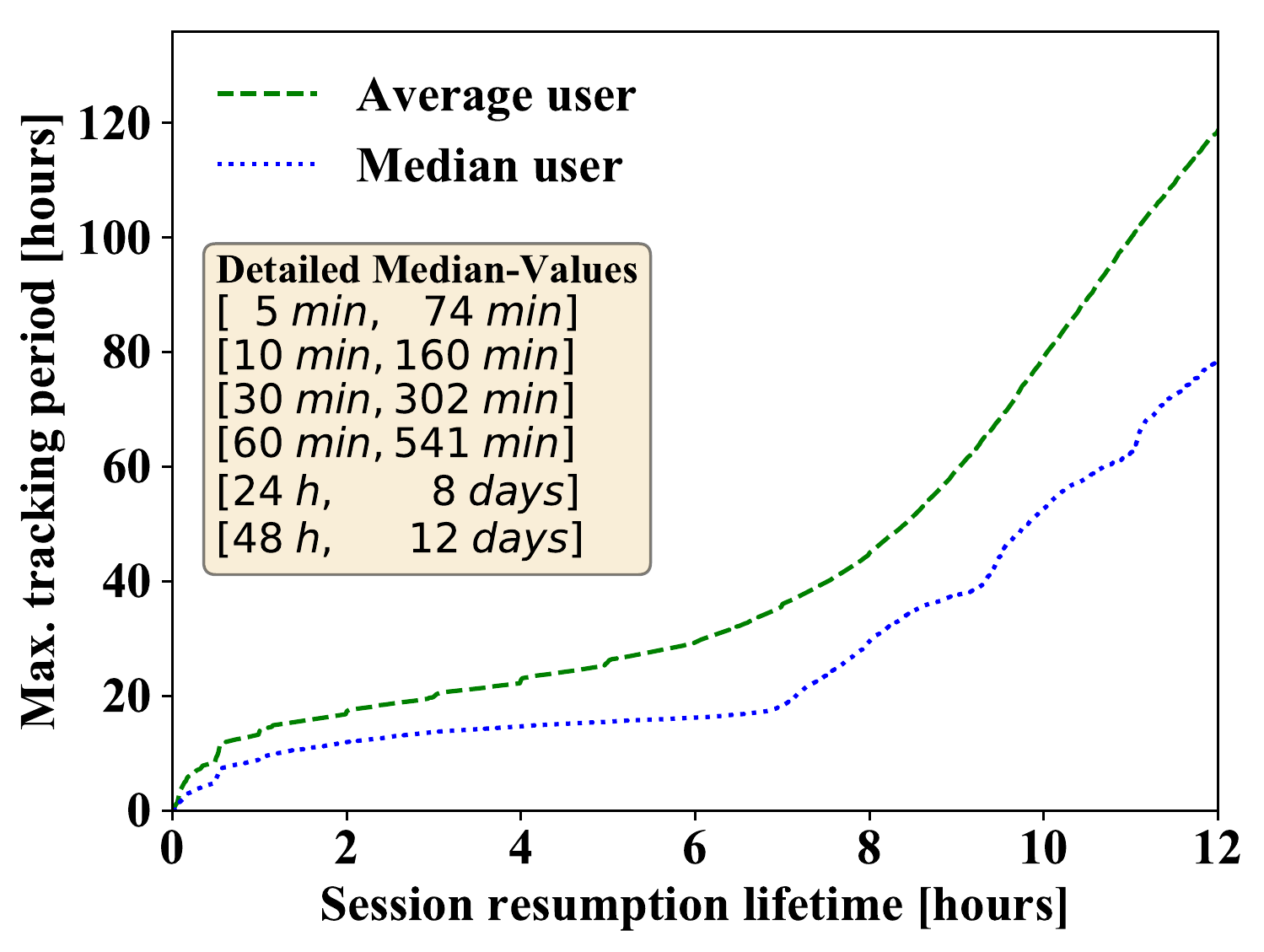}
\caption{The longest consecutive tracking period per user plotted as a function of the session resumption lifetime for the prolongation attack scenario.}
\label{fig:DNS_longest_tracking_period}
\vskip -12pt
\end{figure}

Figure~\ref{fig:DNS_longest_tracking_period} shows the longest consecutive tracking period as a function of the session resumption lifetime determined from said DNS data set with the same assumption as stated above.
Each data point depicts the maximum tracking period overall websites of the respective user, or, more formally,
\[ lctp_{\max}(u, l) \mapsto \max_{s \in S} lctp(v_{u, s}, l),\]
where \(S\) denotes the set of all visited sites and \(v_{u, s}\) the sequence of visits by user \(u\) to site \(s\).
We plotted \(lctp_{\max}\) for the median user (blue, dotted line) and the average user (green, dashed line) of our data set within total 3862 users.

We observe, that the length of the longest consecutive tracking period differs widely among users. While some can be tracked with a session resumption lifetime of 12~hours throughout the whole 61-day period of data collection, the median user can only be tracked for a period of about three days.

We also note that the gradient of these plots exhibits stronger increments for session resumption lifetimes shorter than 20~minutes or longer than seven hours. It seems as if these seven hours were chosen to overcome a user's sleeping phase with reduced online activity because for such longer lifetimes, the tracking period becomes longer than one day.

\subsubsection{Share of Permanently Trackable User}

Next, we evaluate the share of those users in our DNS data set that can be monitored throughout the 61-days sample period. We uphold the assumptions as stated above.

\begin{definition}
We define a user as permanently trackable by a given site, if the visits to this site fulfil the consecutive tracking period property and the time between the first (resp.\ last) visit and the start (resp.\ end) of the sample is less than or equal to the given session resumption lifetime. With the last restriction we ensure, that the user tracking can possibly continue beyond the boundaries of our sample period.
\end{definition}

\begin{figure}
\centering
\includegraphics[width=0.45 \textwidth]{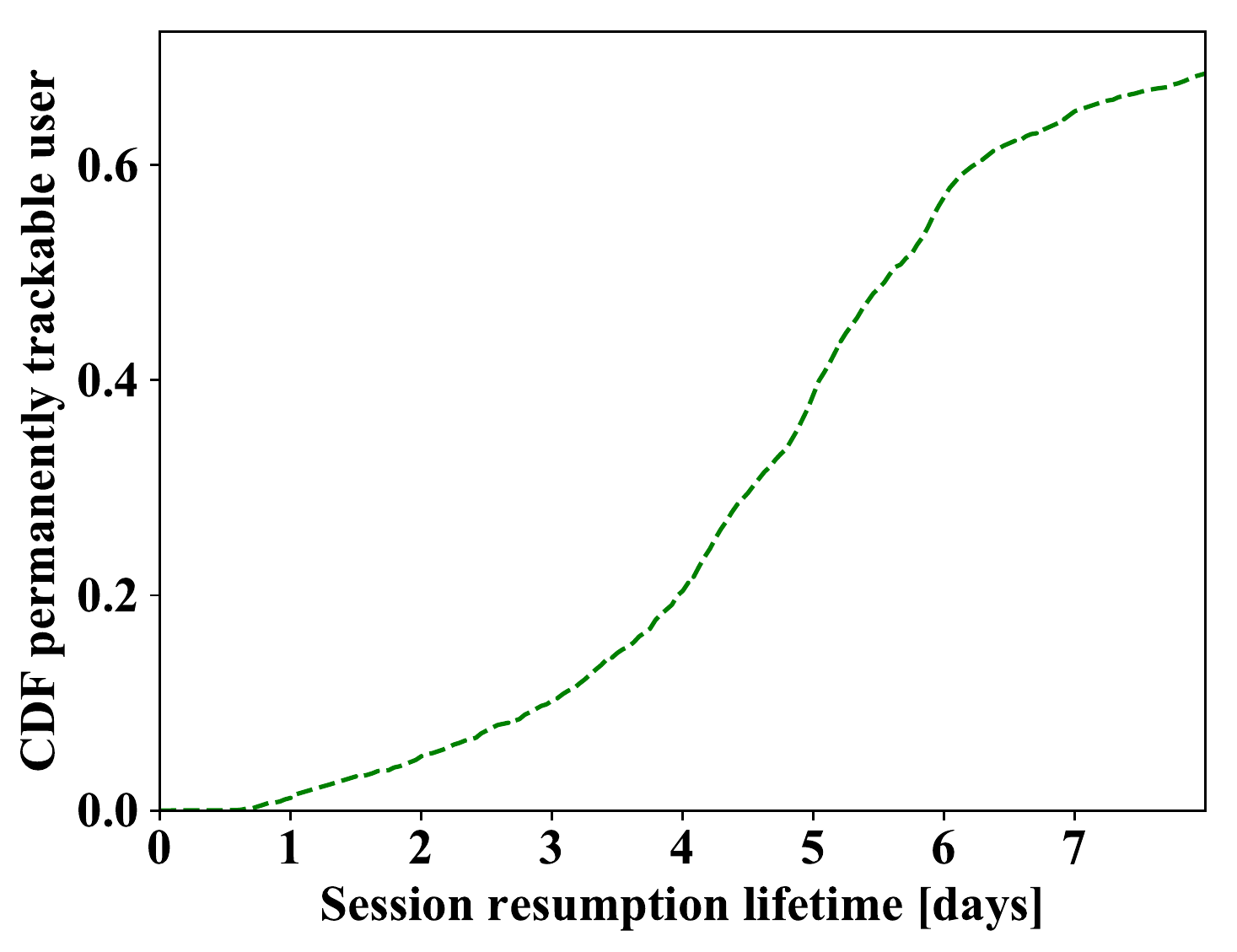}
\caption{Cumulative distribution of permanently trackable user over session resumption lifetime for the prolongation attack scenario.}
\label{fig:DNS_max_tracking_total}
\vskip -12pt
\end{figure}

As can be seen in Figure~\ref{fig:DNS_max_tracking_total}, with a session resumption lifetime of seven days, $65\%$ of all users within our data set can be permanently tracked by at least one website. By limiting the session resumption lifetime to 24~hours, the share of permanently trackable user declines to only $1.3\%$.

\subsubsection{Ratio of Resumed Revisits}

In the interest of providing empirical data to quantify performance gains from resumed session, we investigate the impact of different session resumption lifetimes on the ratio of resumed revisits. We denote revisits which happen within the session resumption lifetime of the previous visit as \textit{resumed revisit}.
\begin{definition}
  We define as \textit{resumption ratio} of a sequence of visits \(v_1,\dots,v_n\) by the same user to an online service given a session resumption lifetime \(l\) as
  \[ rr(v, l) \mapsto \frac{|\{i \in \{1,\dots,n-1\} \wedge |t_{i} - t_{i+1}| \le l\}|}{n - 1}, \]
  where \(t_i\) denotes the time of visit \(v_i\).
\end{definition}

\begin{figure}
\centering
\includegraphics[width=0.45 \textwidth]{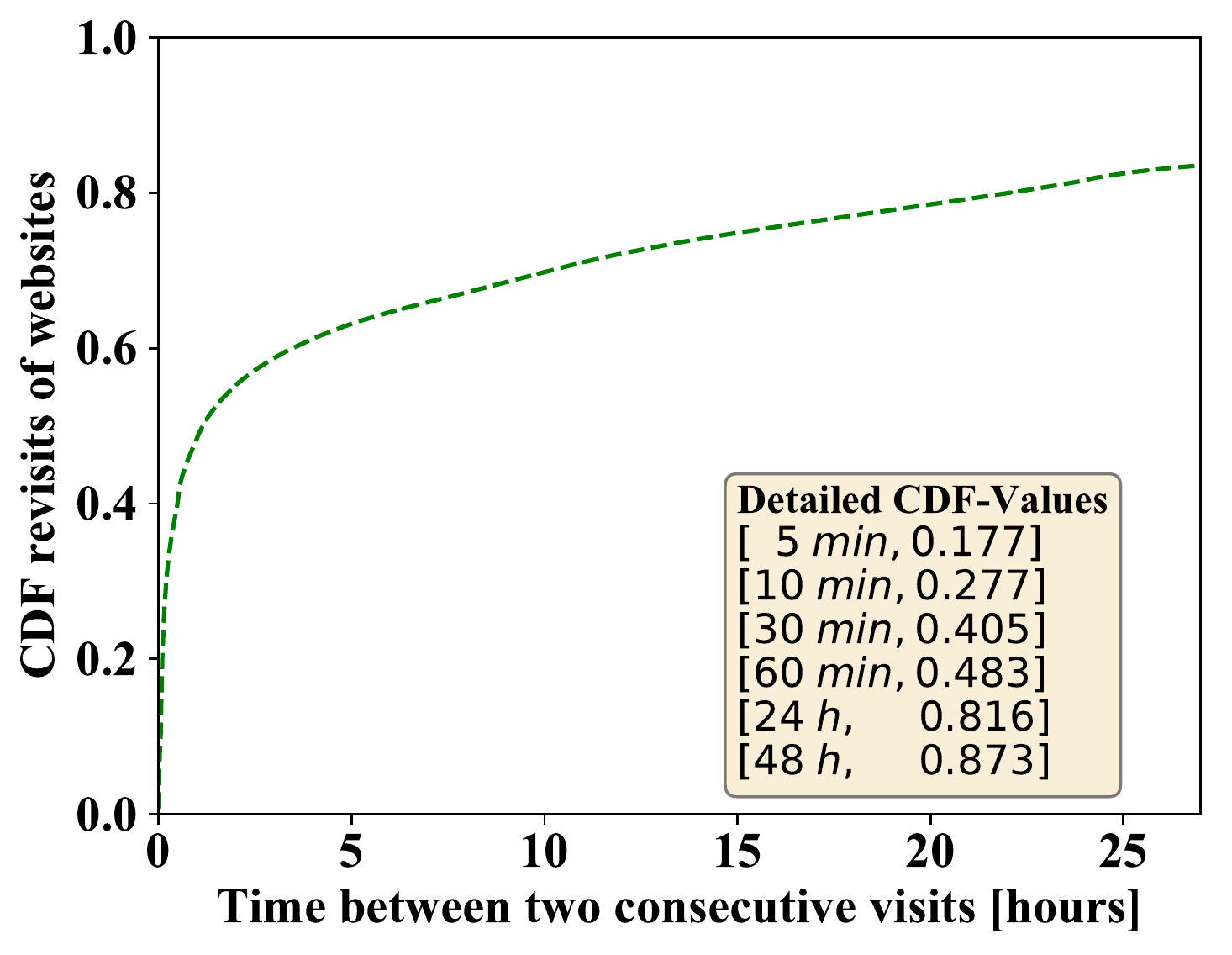}
\caption{Cumulative distribution of revisits of websites plotted over the time between two visits by the same user.}
\label{fig:DNS_Cumulative}
\vskip -12pt
\end{figure}

Figure~\ref{fig:DNS_Cumulative} shows the cumulative distribution of website revisits as a function of the interval between two visits by the same user.

We observe a large share of $17.7\%$ of all revisits can be resumed with a session resumption lifetime of five minutes.
It can be further observed, that the probability of a revisit decreases continuously with the time since the last visit.
Moreover, we find that about half of all revisits occur during the first hour and $95.2\%$ within the first week.

\section{Countermeasures}\label{sec:countermeasures}

Ultimately, this leads us to a discussion of potential countermeasures.
\textbf{A complete protection against tracking via TLS session resumption is achieved by deactivating this feature} as it is practised by the privacy-friendly JonDoBrowser and Tor Browser. A strict deactivation consequently excludes a browser from performance gains, such as the reduced number of round trips of TLS~1.3 0-RTT session resumption.

RFCs on session resumption mechanisms such as TLS~1.3~\cite{TLS13} should be extended in a way, that they exclude a possible prolongation of the lifetime of session resumption mechanisms (see Section~\ref{sec:lifetime}).
We recommend that a server-initiated renewal of a session identifier must not lead to a prolongation of client-side expiry dates. Instead, \textbf{a client must stick to the expiry date of the initial session identifier}. This protects against the prolongation attack (see Section~\ref{sec:lifetime}).

The \textbf{recommended upper limit of the  session resumption lifetime in TLS~1.3~\cite{TLS13} of seven days should be reduced} to hinder tracking based on this mechanism.
We propose an upper lifetime limit of ten minutes based on our empirical observations.
We note, that more than $80\%$ of the Alexa Top Million Sites restrict the session resumption lifetime to less or equal to ten minutes by their own choice and $27,7\%$ of all revisits of a site occur during this period.
Furthermore, the average visit duration of popular websites is of the order of ten minutes~\cite{similarweb}, thus this lifetime limit hinders the correlation of multiple page visits by the same user.

\textbf{Browser vendors should address the issue of third-party tracking via TLS session resumption}, either by deactivating session resumption for third-parties or by allowing only session resumptions to third-parties if the first party site is identical.

Furthermore, considering that TLS~1.3 1-RTT session resumption does not lead to a reduced number of round trips compared to the full handshake, the temporal gains of this mechanism have its origin in the reduced computational complexity of the abbreviated handshake and are rather small. For a latency of 133~ms, which is in the range of 3G mobile network connections, we approximate the temporal gain of TLS~1.3 1-RTT to be in an order of $1\%$ compared to the full handshake. Thus, \textbf{due to the low temporal gains, it seems acceptable to deactivate TLS~1.3 1-RTT session resumption} for privacy reasons.

TLS~1.3 0-RTT provides higher temporal gains by reducing the number of required round trips. However, it is not a replacement of TLS~1.3 1-RTT due to its reduced security guarantees (see Section ~\ref{sec:comparison}). Thus, by \textbf{limiting the support for session resumption to TLS~1.3 0-RTT} the number of resumed sessions should decrease, while the temporal gains are rather high.

Finally, we reported in Section~\ref{sec:server_security} that at least $10\%$ of the Alexa Top Million Sites use their STEK for a period of at least two weeks. Since the STEK makes it possible to decrypt all sessions that use session tickets from such a site, we suggest that \textbf{a minimum STEK change rate should be standardised} to reduce this vulnerable period.

\section{Related Work}\label{sec:relwork}

The impact of TLS on the privacy of its users has already been a focus of previous research. Wachs et al.~\cite{wachs2017push} show in their work, that the TLS \textit{client certificate authentication} transmits unique client certificates in plaintext and thus allows a passive eavesdropper to re-identify and track users. However, websites do not widely use TLS \textit{client certificate authentication} due to its complexity~\cite{wachs2017push}. Thus, it is less feasible to be employed by online tracking service to observe users' browsing behaviour.

Empirical research by Hus\'{a}k et al.~\cite{husak2016https} investigated the feasibility of monitoring TLS handshakes to fingerprint and identify clients. They found, that especially the supported cipher suite lists vary among various client applications and their versions.
This allows them, to infer the client application with a certain precision based on the observed cipher suit list.
While this result may be beneficial to network security monitoring to detect anomalies, it is not suitable for commercial user tracking because of the few observed TLS client configurations during the handshake, do not allow to uniquely distinguish users.

Springall et al.~\cite{springall2016measuring} investigated security problems of TLS session resumption such as STEK lifetime, forward secrecy and TLS state sharing based on measurements of the Alexa Top Million Sites. 

However, to the best of our knowledge, the feasibility of user tracking based on TLS session resumption has not been investigated so far.
The privacy implications of session resumption have only been of concern to software projects. The Tor Browser disabled session resumption due to privacy considerations~\cite{TOR}. Moreover, the chromium project lists session resumption mechanisms as capable to allow client identification~\cite{chromium}.

\section{Conclusion}\label{sec:conclusion}

In this paper, we studied the feasibility of user tracking via TLS session resumption. For that, we evaluated the configuration of popular browsers and online services as well as behavioural user patterns. 

Our results indicate, that most major browsers support TLS session resumption. Only three privacy-friendly browsers deactivated this feature. 
Almost all investigated browsers support tracking periods of at least 30~minutes based on TLS session resumption. We even observed several browsers with session resumption lifetimes of at least 24~hours, e.g., Firefox or Safari. 
Additionally, we investigated whether browsers protect against third-party tracking based on TLS session resumption, which allows to re-identify users across different websites in which the third-parties are embedded as well. Our results indicate that the majority of tested browsers in their standard configuration does not protect users against such third-party tracking.

In addition to the browsers, we also checked the TLS configurations of web servers delivering the Alexa Top Million Sites. We found that the majority of these websites use resumption lifetimes of up to ten minutes, which might be an indication that tracking based on TLS session resumption is not widely applied yet. However, we also observe that especially big players like Google and Facebook use exceptionally long session resumption lifetimes of 28 and 48~hours, respectively. Nevertheless, as longer session resumption lifetimes decrease the load on web servers significantly, this is not a clear indication that the tracking technique is used by them.

As the main contribution of this paper, we present a \textit{prolongation attack} against the TLS standard, which allows to extend the tracking period beyond the session resumption lifetime. Based on a real-world data set on DNS traffic from 2010 with 3862~users we found that with a session resumption lifetime of 24~hours there is one website in the dataset that would be able to track the average user over a period of eight days. 
The  draft of TLS~1.3~\cite{TLS13} proposes an upper session resumption lifetime of seven days. We analysed that such a configuration of session resumption mechanisms would make $65\%$ of all users permanently trackable by at least one single website from our data set. 
However, compared to today the data set contains fewer web requests coming from mobile devices. They render tracking based on TLS session resumption easier, because of their always on-property and as they amplified the frequency users access certain (major) websites.

To mitigate the presented privacy problems, we propose countermeasures to the TLS standard and to browser vendors. The most effective technique is to disable TLS session resumption in browsers completely.

In summary, we hope that our work leads to greater awareness of the privacy risks coming from TLS session resumption and fosters further research on this topic.



\begin{acks}
Part of this research has been conducted in the project AppPETs, which is partly funded by the \grantsponsor{BMBF}{German Federal Ministry of Education and Research}{https://www.bmbf.de/en/} under the reference number~\grantnum{BMBF}{16KIS0381K}.
\end{acks}

\bibliographystyle{ACM-Reference-Format}
\bibliography{sample-bibliography}

\end{document}